\documentclass[sigconf]{acmart}
\AtBeginDocument{%
  }

\usepackage{makecell}

\copyrightyear{2026}
\acmYear{2026}
\setcopyright{cc}
\setcctype{by}
\acmConference[KDD '26]{Proceedings of the 32nd ACM SIGKDD Conference on Knowledge Discovery and Data Mining V.2}{August 09--13, 2026}{Jeju Island, Republic of Korea}
\acmBooktitle{Proceedings of the 32nd ACM SIGKDD Conference on Knowledge Discovery and Data Mining V.2 (KDD '26), August 09--13, 2026, Jeju Island, Republic of Korea}
\acmDOI{10.1145/3770855.3817890}
\acmISBN{979-8-4007-2259-2/2026/08}

\usepackage[textwidth=2.25cm,textsize=footnotesize]{todonotes}
\usepackage{multicol}
\usepackage{multirow}
\usepackage{caption}
\usepackage{subcaption}
\usepackage{tabularx}
\usepackage{tikz}
\usepackage{graphicx}
\usepackage{enumitem}

\definecolor{mygreen}{rgb}{0.23, 0.5, 0}

\definecolor{mplGreen}{rgb}{0.0, 0.5, 0}
\definecolor{mplRed}{rgb}{0.84, 0.15, 0.16}

\newcommand{\hlgreen}[1]{\textcolor{mplGreen}{#1}} 
\newcommand{\hlred}[1]{\textcolor{mplRed}{#1}}

\newlength{\heatmapheight}
\begin{document}

\title{Bradley-Terry Rankings for Recommender Systems Across Dataset Taxonomies}


\author{Ekaterina Grishina}
\authornote{Equal contribution}
\email{ergrishina@edu.hse.ru}
\orcid{0009-0004-7060-0391}
\affiliation{%
  \institution{HSE University}
  \city{Moscow}
  \country{Russian Federation}
}

\author{Stepan Kuznetsov}
\authornotemark[1]
\email{stankuznetsov@edu.hse.ru}
\orcid{0009-0002-4555-0324}
\affiliation{%
  \institution{HSE University}
  \city{Moscow}
  \country{Russian Federation}
}

\author{Askar Tsyganov}
\authornotemark[1]
\email{atsyganov@hse.ru}
\orcid{0009-0003-2372-8416}
\affiliation{%
  \institution{HSE University}
  \city{Moscow}
  \country{Russian Federation}
}

\author{Ilya Ivanov}
\authornotemark[1]
\email{ieivanov@hse.ru}
\orcid{0009-0000-5522-5491}
\affiliation{%
  \institution{HSE University}
  \city{Moscow}
  \country{Russian Federation}
}

\author{Daria Korovaitceva}
\authornotemark[1]
\email{dkorovaitseva@hse.ru}
\orcid{0009-0002-2474-3821}
\affiliation{\country{}}
\affiliation{%
  \institution{HSE University}
  \city{Moscow}
  \country{Russian Federation}
}

\author{Margarita Rusanova}
\authornotemark[1]
\email{mirusanova@hse.ru}
\orcid{0009-0001-4410-7606}
\affiliation{%
  \institution{HSE University}
  \city{Moscow}
  \country{Russian Federation}
}

\author{Uliana Parkina}
\authornotemark[1]
\email{urparkina@edu.hse.ru}
\orcid{0009-0001-4395-3405}
\affiliation{%
  \institution{HSE University}
  \city{Moscow}
  \country{Russian Federation}
}

\author{Alexander Derevyagin}
\authornotemark[1]
\email{aaderevyagin\_1@edu.hse.ru}
\orcid{0009-0009-4563-4158}
\affiliation{\country{}}
\affiliation{%
  \institution{AXXX}
  \institution{HSE University}
  \city{Moscow}
  \country{Russian Federation}
}

\author{Evgeny Frolov}
\email{evgeny.frolov@outlook.com}
\orcid{0000-0003-3679-5311}
\affiliation{%
  \institution{AXXX}
  \institution{HSE University}
  \city{Moscow}
  \country{Russian Federation}
}

\author{Sergey Samsonov}
\email{svsamsonov@hse.ru}
\orcid{0000-0002-0203-2028}
\affiliation{%
  \institution{HSE University}
  \city{Moscow}
  \country{Russian Federation}
}

\author{Anton Lysenko}
\email{av.lysenko@hse.ru}
\orcid{0000-0002-9369-7104}
\affiliation{%
  \institution{HSE University}
  \city{Moscow}
  \country{Russian Federation}
}

\renewcommand{\shortauthors}{Grishina et al.}

\begin{abstract}
    The ranking of recommendation algorithms is a challenging problem since model performance is sensitive to dataset characteristics such as sparsity, sequential structure, and scale. This drives a demand for a proper methodology for fair comparison between algorithms. Naive aggregation of performance metrics (e.g., averaging NDCG over benchmarks) can yield misleading rankings, undermining practical selection. To address this problem, we introduce a novel, data-driven ranking methodology based on Bradley-Terry (BT) model. We demonstrate that the obtained ranking depends on key dataset statistics. Additionally, we propose a novel metric for evaluating ranking consistency and demonstrate robustness of our ranking to incomplete data. Finally, we introduce a dataset-specific methodology for ranking algorithms on unseen datasets without running the models, relying on extensions of the Bradley–Terry framework, including BT trees and BT models with covariates.

\end{abstract}

\ccsdesc[500]{Information systems~Recommender systems}
\ccsdesc[500]{Mathematics of computing~Probabilistic algorithms}
\ccsdesc[500]{Computing methodologies~Learning to rank}

\keywords{Bradley-Terry Model, Pairwise Comparison, Recommender Systems, Algorithm Ranking}

\received{20 February 2007}
\received[revised]{12 March 2009}
\received[accepted]{5 June 2009}

\maketitle

\newcommand\kddavailabilityurl{https://doi.org/10.5281/zenodo.20383718}
\ifdefempty{\kddavailabilityurl}{}{
    \begingroup
    \small
    \noindent
    \raggedright
    \textbf{Resource Availability:}\\
    The source code of this paper has been made publicly available at \url{\kddavailabilityurl} or \url{https://github.com/fallnlove/btl_recsys}.
    \endgroup
}

\section{Introduction}

In recent years, the field of recommendation systems has been characterized by a wide variety of tasks, datasets, and algorithmic approaches. Almost every dataset has its own specifics: the types of interactions, sparsity, scale and time structure can vary. As a result, the same algorithm can demonstrate high performance on some subset of datasets and significantly lose on others. This makes it challenging for practitioners to choose the most suitable method for a specific task without costly direct comparison. From an academic research perspective, the variety of existing algorithms poses a challenge of robust and scalable evaluation.  Direct comparison by individual metrics and datasets often gives a fragmentary picture and scales poorly with an increasing number of methods and datasets. These issues highlight the need for a principled methodology for fair algorithm comparison and dataset-aware selection. 

In this work, we propose to evaluate the recommendation algorithms within the probabilistic ranking framework based on the Bradley-Terry model \cite{bradley-terry} and its modifications, including covariate-adjusted \cite{schauberger2019btllasso, fan2024uncertainty} and tree-based variants \cite{zeileis2008modelpartitioning, gijsbers2024amlb, strobl2011accounting, zeileis2011psychotree}. We conduct an extensive benchmark of 14 recommender algorithms across 89 datasets, using a framework that aggregates disparate experimental results into a single quality scale to provide more informed recommendations for algorithm selection based on dataset taxonomy. Our main contributions are:

\begin{itemize}[noitemsep,topsep=0pt,leftmargin=1em]
\item We introduce a theoretically grounded methodology for aggregating repeated evaluations across diverse datasets into a probabilistic ranking of recommendation algorithms based on the Bradley--Terry model. This yields rankings tailored to key dataset classes, such as sequential or sparse datasets, supporting baseline selection (Section \ref{sec:ranking_classes}).
\item We introduce the "transitive triplets" metric for evaluating ranking consistency under incomplete benchmark results (Section \ref{sec:validity}). Using this metric, we show that Bradley--Terry-based rankings are substantially more stable than rankings obtained by simple metric aggregation (Section \ref{sec:comparison}).
\item  We extend the framework with dataset-aware baseline selection using BT trees and covariate-adjusted BT \cite{schauberger2019btllasso, strobl2011accounting} enabling ranking prediction on a target dataset without running candidate models (Section \ref{sec:bt_covariate_eval}).
\end{itemize}




\section{Related work}

Pairwise comparison models have a long history as a principled approach to ranking, originating with the Bradley-Terry (BT) model \cite{bradley-terry}, first formulated by Zermelo \cite{zermelo1929berechnung}. Due to their interpretability properties, BT-based models have been widely adopted in various domains, including sports, chess, and other competitions, ranking of AI models \cite{chiang2024chatbot}. Several extensions of the Bradley-Terry model have been proposed to address different comparison settings, such as the Plackett-Luce model \cite{plackett1975analysis}, which builds rankings based on listwise comparisons instead of pairwise. Other extensions include the Thurstone-Mosteller \cite{thurstone1927tm, mosteller1951tm} and Rao-Kupper \cite{rao1967ties} models, which account for ties or latent score variability. Recently, Bayesian formulations of the Bradley-Terry model have been proposed, allowing for uncertainty-aware inference and more robust pairwise comparisons. These formulations have been applied to evaluation of classification algorithms across multiple datasets \cite{wainer2023bayesianbt}. A notable extension of the Bradley-Terry model is the covariate-adjusted BT \cite{schauberger2019btllasso, fan2024uncertainty}, that allows incorporating information about  the varying context of each singular comparison (e.g. dataset characteristics). A special case of this approach is BT trees \cite{strobl2011accounting}, which recursively partition the subjects and identify groups of subjects with homogeneous preference scalings in a data-driven way. For instance, in AutoML benchmark \cite{gijsbers2024amlb}, BT-trees have been used to   discover subsets of tasks where the relative AutoML framework rankings differ.
 
Beyond statistical models, a parallel line of work focuses on algorithmic methods for reconstructing rankings from pairwise comparisons. These approaches aim to efficiently recover approximate or exact rankings via dynamic or static algorithms \cite{heckel2018approxranking, wauthier2013efficient}, attempting to make as little pairwise comparisons as possible. It is important to note that such methods essentially presume the existence of certain true ranking or suppose the possibility of comparing items dynamically. Thus, they are either inapplicable in real-life scenarios or the resulting rankings may depend on the comparison schedule and lack uncertainty quantification.


Regarding the comparison of algorithms over multiple datasets, traditional methods rely on frequentist statistical tests \cite{demsar2006tests} or aggregated performance metrics \cite{shevchenko2024benchmarking}. Although these approaches can provide global rankings, they may be unstable or misleading due to the weak theoretical foundation. Bayesian alternatives have been proposed to address these shortcomings by enabling direct pairwise comparisons and probabilistic statements about algorithm performance \cite{benavoli2016bayesiancomparison}. Recent works apply Bradley-Terry-type models to algorithm benchmarking, demonstrating their suitability for deriving consistent and interpretable rankings across datasets \cite{wainer2023bayesianbt}. However, to the best of our knowledge, such approaches have not yet been applied to RecSys algorithms.






\section{Bradley-Terry model}
\label{sec:bradley-terry-model}

Our approach is based on probabilistic ranking. In this section, we first describe the pairwise comparison model; then we move on to the methods for computing the ranking from pairwise comparisons. Finally, we describe the Plackett-Luce model, which produces rankings based on multiple (listwise) comparisons. 

\subsection{Pairwise comparisons properties}
The Bradley-Terry model \cite{zermelo1929berechnung, bradley-terry} is a probability model for ranking "players" in "tournaments" based on pairwise comparisons or "matches". Consider the setting of a tournament between $n$ players. Within this model, the probability that the player $i$ wins over player $j$ is modeled as 
\[
\mathrm{Pr}(i\succ j)=\frac{p_i}{p_i+p_j},
\]
where $p_i\geq 0$ are weights assigned to each of $n$ players, representing their strengths or abilities. The final ranking of the players is defined by the rank of their weights $p_i$. The weights are invariant to a multiplicative constant, so in order to get a unique set, an additional constraint is usually imposed, e.g., $\sum_{i=1}^n p_i=1$ or $\prod_{i=1}^n p_i=1$.

In our setup, instead of "players", we want to rank recommender algorithms based on pairwise comparisons between them. We say that the algorithm $i$ beats the algorithm $j$ on a dataset, if it achieves a higher metric on this dataset. This is how we get the tournament table $W$ with $W_{ij}$ being the number of datasets, where algorithm $i$ beats algorithm $j$. We assume that $W_{ii}=0$. 

If the algorithms have achieved almost the same metric value, we can classify it as a ``tie''. There are many approaches to handling ties, e.g \cite{davidson1970extending} modifies original model by introducing additional parameters, while \cite{wainer2023bayesianbt} leaves the original model unchanged, but modifies $W$. Based on the work \cite{wainer2023bayesianbt}, we take the ties into account by adding 0.5 to both $W_{ij}$ and $W_{ji}$ in case if the achieved metric intervals $\mathrm{metric_i\pm std_i}$ and $\mathrm{metric_j\pm std_j}$ overlap (see definition of intervals in Section \ref{sec:exp_setup}). 
After that, we run any standard (unaware of ties) Bradley-Terry algorithm on the obtained matrix $W$.

\subsection{Estimating parameters in the Bradley-Terry model \label{sec:bt_estimation}}

Within the classic Bradley-Terry (BT) model \cite{bradley-terry, zermelo1929berechnung}, the likelihood of the tournaments' outcomes is $\prod_{1 \leq i,j \leq n}[P(i\succ j)]^{W_{ij}}$ and the loglikelihood is
\[
\begin{split} 
\ell(p) 
&= \ln \prod_{1 \leq i,j \leq n}[\mathrm{Pr}(i\succ j)]^{W_{ij}} =\\ 
&=\sum_{1 \leq i,j \leq n}[W_{ij}(\ln(p_i)-\ln(p_i+p_j))].
\end{split}
\]
Zermelo \cite{zermelo1929berechnung} showed that this expression has a single maximum and suggested to find it by a simple iteration:
\[
p'_i \leftarrow \frac{\sum_{j=1}^{n} W_{ij}}{\sum_{j=1}^{n}(W_{ij} +W_{ji})/(p_i+p_j)},\]
\[p_i\leftarrow p'_i/\left(\prod_{j=1}^np_j\right)^{1/n}.
\]
Theoretical properties of the Bradley-Terry model estimates are fairly well understood, see recent works \cite{gao2023uncertainty}, \cite{spokoiny2025semiparametric}, and references therein.
\par 
However, the classic Bradley-Terry model has a limitation: it provides only a point estimate of parameters $p_i$ without confidence intervals. To address it, the authors  of \cite{wainer2023bayesianbt} proposed the Bayesian Bradley-Terry model.  This model assumes that there are $N_{ij}=N_{ji}$ comparisons between algorithms ($N_{ij}=W_{ij}+W_{ji}$ if there are no ties). Using natural logarithms of parameters $\beta_i=\ln p_i$, the authors introduce the following model:
\begin{equation}
\label{eq:bayes-bt}
\begin{split}
W_{ij} &\sim \mathrm{Binomial}\left(N_{ij}, \frac{e^{\beta_i}}{e^{\beta_i}+e^{\beta_j}}\right)\,, \\
\beta_i &\sim \mathrm{Normal}(0, \bar{\sigma}), \\
\bar{\sigma} &\sim \mathrm{LogNormal}(0, 0.5).
\end{split}
\end{equation}
The parameters of this model are estimated using Metropolis-Hastings. Metropolis-Hastings outputs a set of samples for the parameters $\beta_{i}, \bar{\sigma}$, and the final ranking is determined by the mean $\beta_i$ across all samples. The samples from MCMC allow to build confidence intervals for the weights $p_i$ and probabilities $P(i\succ j)$.

Another approach called \textit{rank centrality} or spectral method \cite{oh2017rank, gao2023uncertainty} models the comparisons with a random walk on the underlying Erdős-Rényi graph with adjacency matrix $A$. It models the Markov chain with $n$ states, given the sample transition matrix $P$: 
\[
P_{ij} \equiv 
\begin{cases} 
\frac{1}{2nd} A_{ij} \bar{w}_{ji}, & i \neq j, \\
1 - \frac{1}{2nd} \sum_{k: k \neq i} A_{ik} \bar{w}_{ki}, & i = j.
\end{cases}
\]
where $\bar{w}_{ij}=W_{ij}/(W_{ij}+W_{ji})$ is the ratio of wins and $d$ is the edge density of the graph with adjacency matrix $A$. This transition matrix is modeled with
\[
P^*_{ij} \equiv \mathbb{E}(P_{ij} | A) =
\begin{cases} 
\frac{1}{2nd} A_{ij} \sigma(\theta_j^* - \theta_i^*), & i \neq j, \\
1 - \frac{1}{2nd} \sum_{k: k \neq i} A_{ik} \sigma(\theta_k^* - \theta_i^*), & i = j,
\end{cases}
\]
where $\sigma$ is the sigmoid function and $\theta^\ast_i$ are the parameters (Bradley-Terry weights). This transition matrix admits the stationary measure
\[
\pi^* \equiv \left( \frac{e^{\theta_1^*}}{\sum_{k=1}^n e^{\theta_k^*}}, \dots, \frac{e^{\theta_n^*}}{\sum_{k=1}^n e^{\theta_k^*}} \right),
\]
and the approach of rank centrality estimates it using the stationary measure
$\hat \pi$ of $P$:
\[
\hat{\pi}^\top P = \hat{\pi}^\top.
\]
Imposing the condition $\sum_i \theta^\ast_i=0$, we get
\[\theta^\ast_i = \log \hat\pi_i - \frac{1}{n}\sum_{k=1}^n \hat \pi_k.\]

\subsection{Multiple comparisons}
Plackett-Luce (PL) model \cite{plackett1975analysis, debreu1960individual} is an extension of Bradley-Terry model to listwise rankings, that is, simultaneous comparisons of more than two objects. Plackett-Luce model takes $R$ rankings 
\[
(y_{i_1} \succ y_{i_2} \succ \dots \succ y_{i_{T_i}})
\]
of $T_i\leq n$ objects as input, where $n$ is the total number of objects, and models probabilities
\[\mathrm{Pr}(y_{i_1}\succ y_{i_2} \succ \dots \succ y_{i_{T_i}})=\prod_{k=1}^{T_i}\frac{p_{i_k}}{\sum_{j=k}^{T_i}p_{i_j}}.\]
The authors of \cite[Section 4]{caron2012efficient} proposed to introduce for  $i=1,\dots, R$ and $j=1, \dots T_i-1$ latent variables $Z=\{z_{ij}\}$: 
\[
f(z|Data, p) = \prod_{i=1}^{R} \prod_{j=1}^{T_i-1} \mathrm{Exp}(z_{ij}; \sum_{k=j}^{T_i} p_{i_k}),
\]
leading to loglikelihood
\[\ell(p, z)=\sum_{i=1}^R\sum_{j=1}^{T_i-1} \log p_{y_{ij}}- \left(\sum_{k=j}^{T_i}p_{y_{ik}}\right) z_{ij}.\]
This loglikelihood can be maximized using EM algorithm. 

An alternative approach proposed by the authors of \cite{caron2012efficient} is to sample from $f(p, z| Data)$ using Gibbs sampler:
\[Z_{ij}^{(t)} \mid Data, \lambda^{(t-1)} \sim \mathcal{E}\left(\sum_{k=j}^{T_i} \lambda_{\rho_{ik}}^{(t-1)}\right),\]
\[
\lambda_k^{(t)} \mid Data, Z^{(t)} \sim \mathcal{G}\left(a + w_k,\; b + \sum_{i=1}^R \sum_{j=1}^{T_i-1} \delta_{ijk} Z_{ij}^{(t)}\right),
\]
where $w_k$ is the number of rankings where the $k
$-th individual is not in the last ranking position and
$ \delta_{ijk}$ is the indicator of the event that individual $k$ receives a rank no better than $j$ in the $i$-th
ranking.

In our setup, we obtain comparisons $(y_{i_1} \succ y_{i_2} \succ \dots \succ y_{i_{T_i}})$ by sorting recommender algorithms on each dataset according to their values of a metric.

\section{Validation of the rankings \label{sec:validity}}
A common metric for assessing the consistency of rankings is rank correlation, such as Kendall’s $\tau$ or Spearman’s $\rho$. To evaluate the validity of a given ranking, one may take the mean rank correlation between that ranking and the per‑dataset rankings. However, when a metric value is missing for a particular algorithm on a given dataset, that algorithm must be excluded from both rankings before computing the correlation. Consequently, standard rank correlation measures do not capture the robustness of a ranking with respect to missing data. To address both validity and robustness simultaneously, we propose a metric based on the number of transitive triplets in the ranking.

The ranking $i_1^d\succ i_2^d \succ i_3^d$ is called transitive on dataset $d$ if $i_1^d$ wins $i_2^d$, $i_2^d$ wins $i_3^d$ and $i_1^d$ wins $i_3^d$. In an ideal ranking all triplets are transitive, i.e., all duels between any three algorithms are consistently ordered. Hence, we can use the ratio of transitive triplets across datasets as a quality metric. We count only distinct triplets, so the total number of ordered triplets is $(^n_3)\cdot D$, where $n$ is the number of algorithms, and $D$ is the number of datasets. If there are missing comparisons in data, then the total number of ordered triplets decreases.. A higher ratio indicates a better ranking:
\[\text{Triplet ratio} = \frac{1}{(^n_3) \cdot D - \mathrm{missing}}\sum_{d=1}^{D}\sum_{i_1^d \not= i_2^d \not= i_3^d}^n I\{i_1^d\succ i_2^d\succ i_3^d\}.\]
In addition, this metric can take ties into account. If the metrics achieved by two algorithms $i$ and $j$ are very close (intervals $\mathrm{metric_i\pm std_i}$ and $\mathrm{metric_j\pm std_j}$ overlap), we consider the comparison a tie. Taking ties into account, we say that the triplet is transitive if $i_1\succeq i_2 \succeq i_3$. For all ranking methods, we rely solely on the weights $p_i$ to rank algorithms and do not treat overlapping confidence intervals produced by statistical models as ties in a ranking.

\section{Experimental Setup \label{sec:exp_setup}}
\subsection{Datasets and recommendation algorithms}
To ensure a broad and representative evaluation, we selected a diverse collection of recommendation datasets covering different domains and structural properties. Most come from the APS benchmark set \cite{vente2025aps}, which provides a collection of standard datasets popular in recent recommender systems studies. To further increase the diversity, we included a few additional datasets.

Following common practice \cite{kang2018sasrec} in recommender system evaluation, all datasets were preprocessed using a 5-core filtering scheme. The final statistics of the datasets after preprocessing are reported in Table \ref{table:dataset_stats_clean} in Appendix. For further analysis, we grouped datasets along four dimensions: density, user-item ratio, mean interactions per user, and sequentiality. The assessment of sequentiality follows the methodology proposed by Klenitskiy et al. \cite{klenitskiy2024sequential} based on 2-gram statistics. In experiments comparing performance across these dimensions, we evaluated the models on the top-20 datasets with the highest and lowest values for each metric (e.g., the 20 most and the 20 least sequential). Datasets without timestamp information were considered non-sequential.

We utilize a set of standard and widely adopted baselines to ensure a comprehensive evaluation. Our selection includes trivial non-personalized heuristics (Random, PopRandom) and three neighborhood-based algorithms (User-KNN, Item-KNN, Seq-KNN). We also examine five matrix factorization and linear models (ALS \cite{hu2008collaborative, takacs2011applications}, BPR \cite{rendle2009bpr}, SGD MF \cite{randle_negative_sampling}, PureSVD~\cite{puresvd}, EASEr \cite{steck2019easer}). Finally, to represent modern developments in the field, we include two graph-based models (LightGCN \cite{lightgcn}, UltraGCN \cite{mao2021ultragcn}) and two advanced sequential architectures: the transformer-based SASRec \cite{kang2018sasrec, mezentsev2024sce} and the tensor-based GASATF \cite{gasatf}. Implementation details are described in Table \ref{table:rec_algorithms} in Appendix.

\subsection{Evaluation protocol}
We follow the experimental setup described in \cite{gusak2025time}, specifically the global temporal split (GTS) protocol. Each dataset is divided into training, validation, and test periods based on global timestamps with a 90/5/5 split ratio. For datasets without timestamps, we assigned random timestamps to conform to this setup.

For validation, we use the last-item strategy, selecting the last interaction during the validation period as the target and for testing, the random-item as a target item was used. For robustness, we average the metrics over 10 different random seeds. Using these results, we obtain metric intervals $\mathrm{mean\_metric_i}\pm \mathrm{std_i}$, which we use to identify ties between algorithms.

We used the Optuna \cite{optuna2019} framework with tree-~structured Parzen Estimator (TPE) sampler to perform a hyperparameter optimization. For all models, hyperparameters are selected by optimizing NDCG@10 on the validation set. We use a single search space for all datasets to ensure fair comparison and set a budget of 200 trials (with 20 startup trials) to balance search depth and cost. Due to the prohibitive computational cost of certain algorithms on larger datasets, the optimization process was occasionally terminated early due to time constraints. In such cases, we selected the best hyperparameters identified up to that point. After selecting the optimal hyperparameters, the model was retrained on the combined training and validation sets, and test metrics were collected.

\section{Experiments}

\subsection{Comparison with aggregation baselines \label{sec:comparison}}


In recommender systems and other domains, it is common to rank algorithms by the mean or sum of metrics across datasets \cite{sertkan2023exploring,qinsidobi}. However, these aggregation methods might yield unstable or contradictory rankings, as the comparison of metrics is usually valid only for a particular problem. Approaches based on the BT model (described in Section \ref{sec:bradley-terry-model}) offer a different perspective by simulating a "competition" between algorithms. In this section, we validate the BT model, while in section \ref{sec:main_bt_covariates} we assess its predictive power and compare with covariate-adjusted extensions.



\paragraph{Quality of rankings.}  To compare the quality of different rankings, we consider the ratio of transitive triplets in a ranking and mean Kendall's $\tau$ between given ranking and per-dataset rankings, as described in Section \ref{sec:validity}. A high-quality ranking is expected to be largely transitive: if algorithm $A$ outperforms $B$, and $B$ outperforms $C$, then $A$ should also outperform $C$. As shown in Table \ref{tab:triangles}, Bradley-Terry  achieves a higher ratio of transitive triplets for all datasets and especially for subsets of datasets with long user history and sparse datasets. Notice that the PL model demonstrates results very similar to the ones of the BT model. We also observe, that the triplet ratio is consistent with Kendall's $\tau$, and by both metrics BT model shows the best results (for more Kendall's $\tau$ results see table \ref{tab:kendall_corr} in Appendix). However, Kendall's $\tau$ can not be used to assess stability of the rankings when there are many missing pairwise comparisons. Instead, we apply our proposed ``triplet ratio`` metric as a proxy to compare the robustness of the rankings.

\begin{table}
\caption{\label{tab:triangles} Ranking quality. Algorithms are compared on all datasets and subsets of sparse and long-history-datasets. Comparison is based on NDCG@10.}
\begin{tabular}{c|cc|cc|cc}
\toprule
& \multicolumn{2}{c}{All} & \multicolumn{2}{c}{Long history} & \multicolumn{2}{c}{Sparse} \\
 & w/o ties & w ties & w/o ties & w ties & w/o ties & w ties \\
\midrule
\multicolumn{7}{c}{Ratio of transitive triplets.}\\
\midrule
Mean & 0.488 & 0.613 & 0.433 & 0.556 & 0.495 & 0.615 \\
Sum & 0.484 & 0.613 & 0.445 & 0.569 & 0.458 & 0.575 \\
BT & \textbf{0.510} & \textbf{0.631} & \textbf{0.471} & \textbf{0.588} & \textbf{0.505} & \textbf{0.624} \\
PL & 0.503 & \textbf{0.631} & 0.450 & 0.576 & 0.486 & 0.614 \\
\midrule
\multicolumn{7}{c}{Mean Kendall's $\tau$.}\\
\midrule
Mean & 0.542 & 0.542 & 0.488 & 0.488 & 0.542 & 0.542 \\
Sum & 0.541 & 0.541 & 0.5 & 0.5 & 0.499 & 0.499 \\
BT & \textbf{0.567} & \textbf{0.564} & \textbf{0.533} & \textbf{0.524} & \textbf{0.565} & \textbf{0.558} \\
PL & 0.558 & 0.558 & 0.502 & 0.502 & 0.547 & 0.547 \\
\bottomrule
\end{tabular}
\end{table}

\paragraph{Stability of rankings.} Since in practice running all compared algorithms on all available datasets may be computationally expensive and time consuming, the results of some pairwise comparisons might be missing. Therefore, it is important to evaluate robustness of difference ranking methods to incomplete data.
 
\begin{figure}
    \centering
        \includegraphics[width=0.8\linewidth]{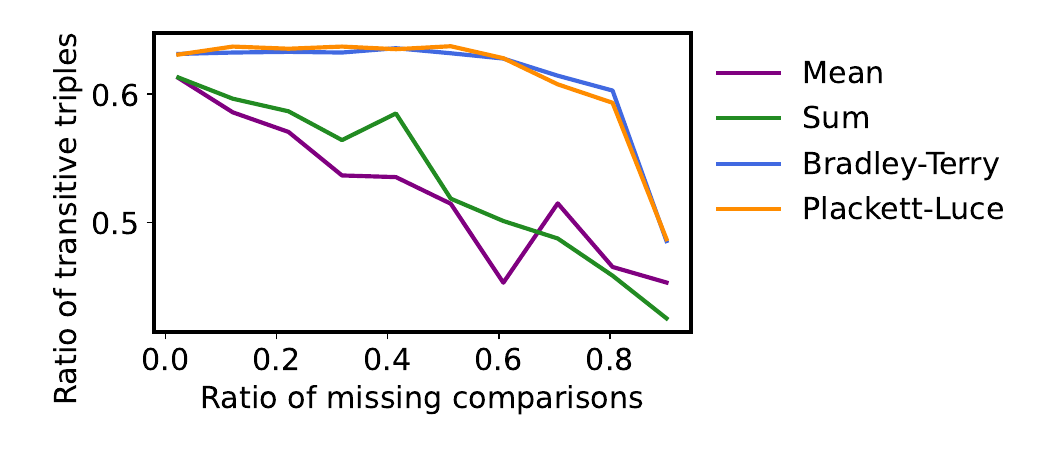}
        \caption{Ratio of transitive triples in rankings of different methods (by NDCG@10) depending on the ratio of missing comparisons in the data.}
        \label{fig:gaps}
        \Description{Line plot shows ratio of transitive triples in rankings of different methods (by NDCG@10) depending on the ratio of missing comparisons in the data. For Bradley-Terry and Placket-Luce, the lines are stable at 0.6 ratio of triples and decline only when ratio of missing values exceeds 0.7. The ratio of triples for Sum and Mean decreases from 0.6 to about 0.4 with increasing ratio of missing comparisons.}
\end{figure}

\begin{figure}
    \centering
    \includegraphics[width=1\linewidth]{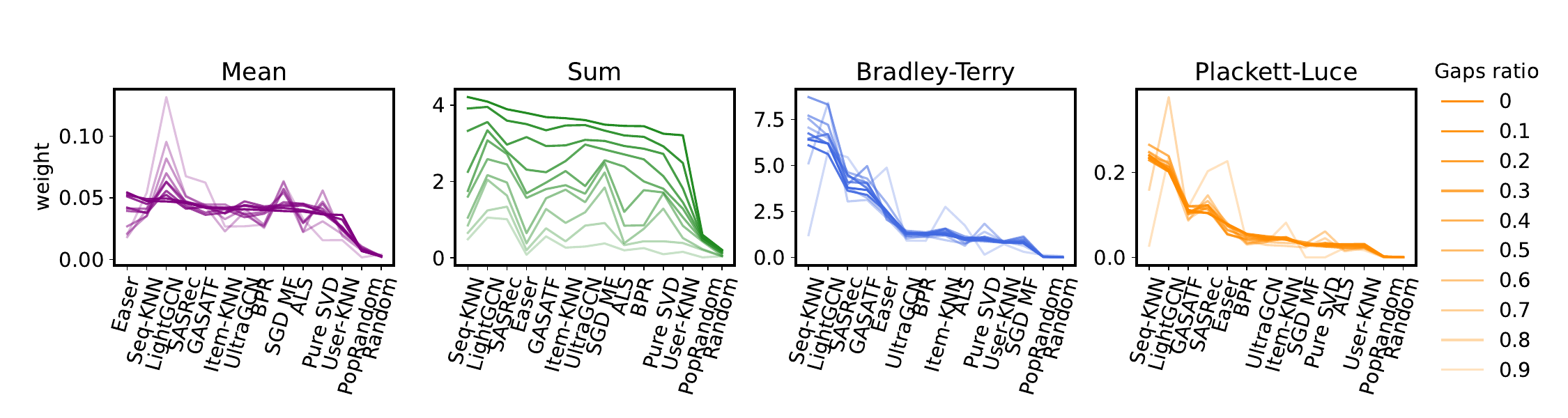}
        \caption{Weights of rankings by NDCG@10 for data with varying ratio of missing values.}
        \label{fig:gaps_weights}
        \Description{Figure shows 4 plots for weights and rankings of Sum, Mean, BT and PL.  When the ratio of missing values increases (shown by fainter color), the weights of Sum and Mean fluctuate, but the weights of BT and PL remain stable.}
\end{figure}

To assess stability, we randomly omit a certain fraction of entries from the table with algorithm's metrics, and recompute the ranking based on the remaining data. Figure \ref{fig:gaps} shows that the ratio of transitive triples in the rankings by  mean and sum of values rapidly decreases as the proportion of missing data increases. In contrast, the ratio of transitive triples remains almost constant for the BT and PL models. While the weights of BT and PL appear stable in Figure \ref{fig:gaps_weights}, the weights of the "sum" and "mean" rankings fluctuate with growing gap ratio. This experiment indicates that BT and PL rankings are substantially more robust to missing data. 

\subsection{Comparison of statistical models \label{sec:different_bt}}
\begin{figure}
    \centering
    \begin{subfigure}[t]{0.47\textwidth}
    \centering
    \includegraphics[width=1\linewidth]{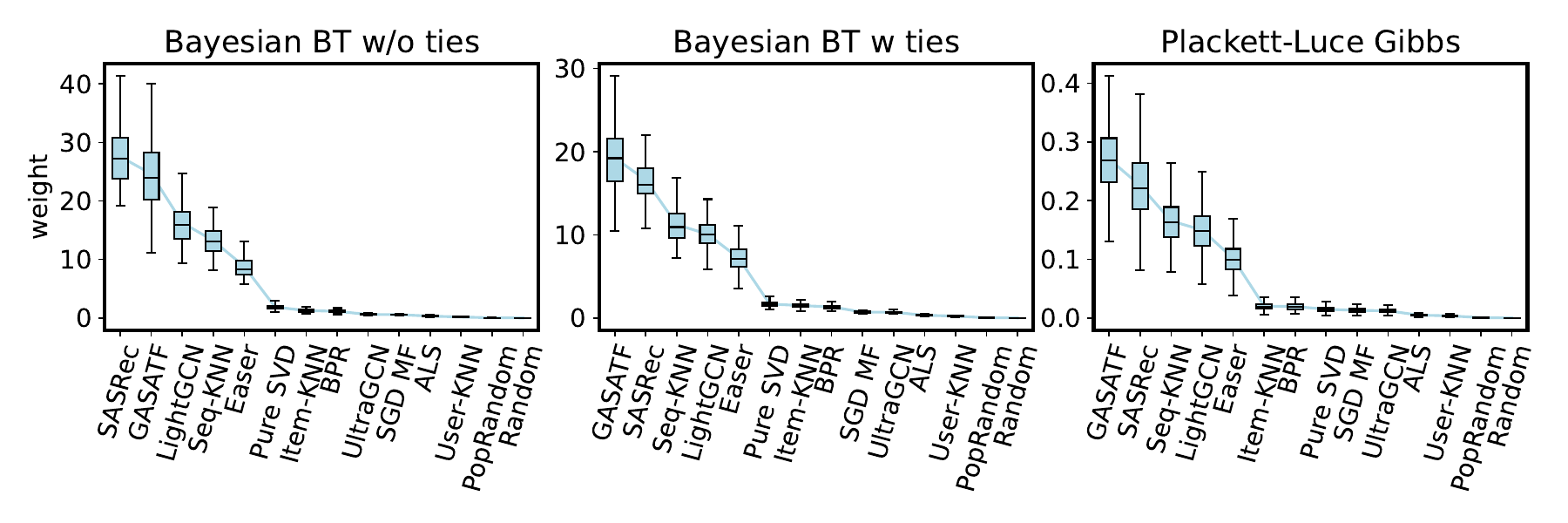}
    \caption{All datasets.}
    \end{subfigure}%
    \vfill
    \begin{subfigure}[t]{0.47\textwidth}
    \centering
    \includegraphics[width=1\linewidth]{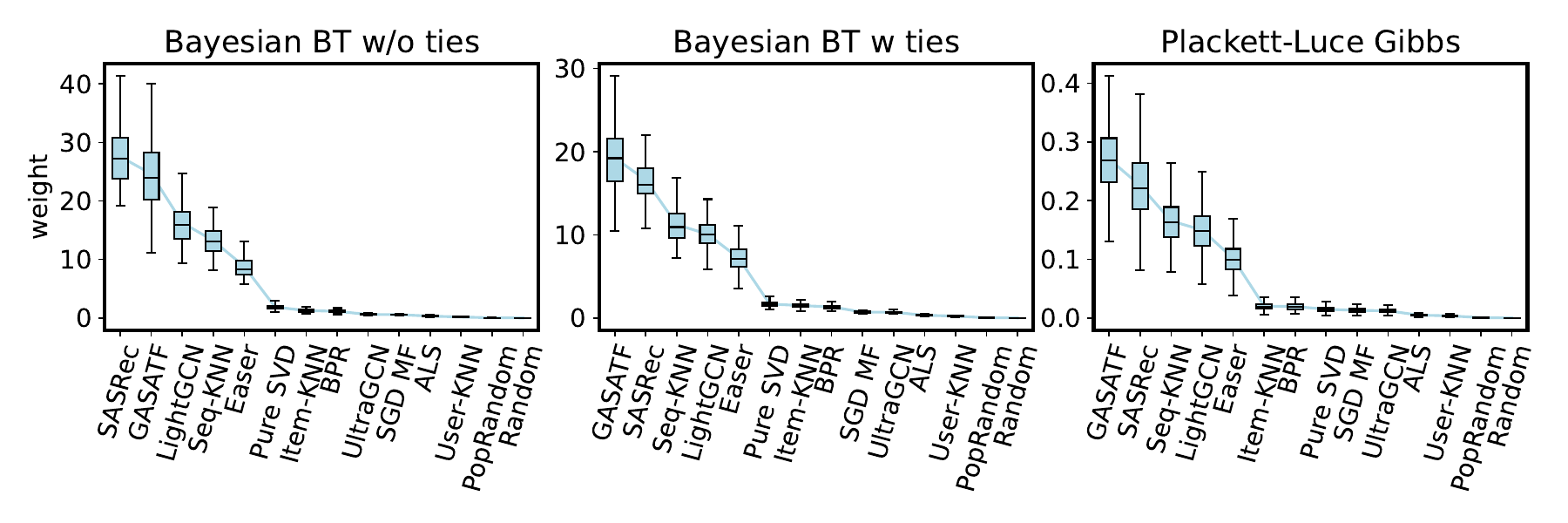}
    \caption{Sequential datasets.}
    \end{subfigure}%
    \vfill
    \begin{subfigure}[t]{0.47\textwidth}
    \centering
    \includegraphics[width=1\linewidth]{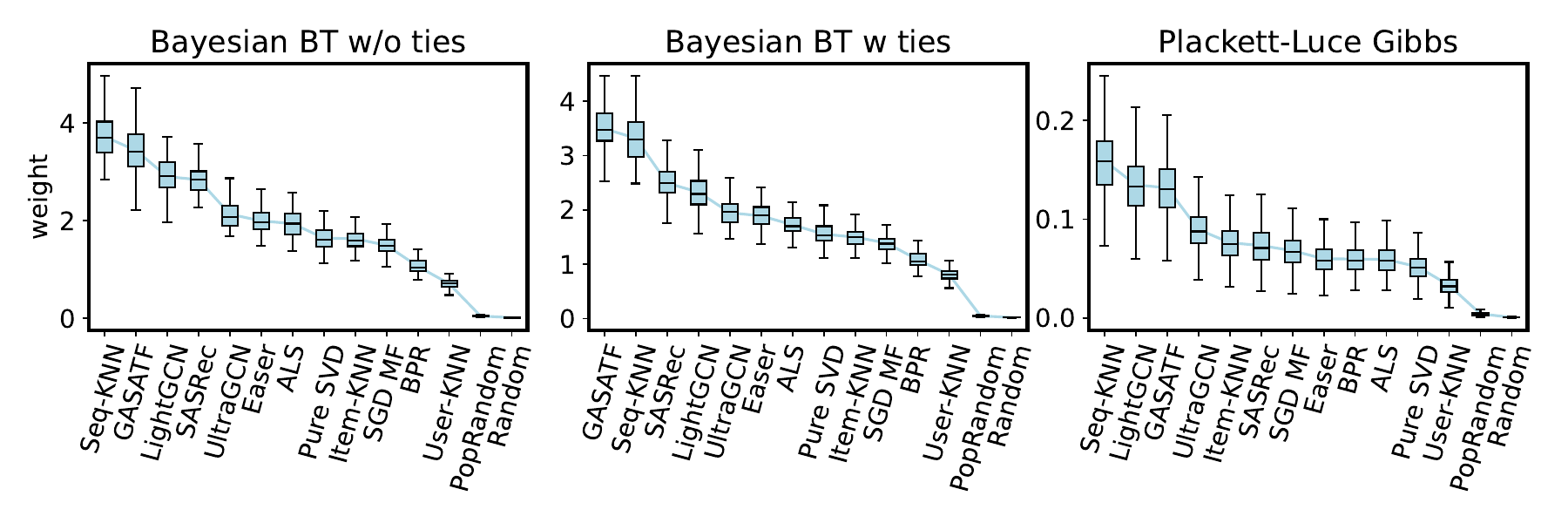}
    \caption{Top 20 datasets with long user history.}
    \end{subfigure}%
    \caption{\label{fig:btl_comparison} Comparison of weights of different models. Metric for rankings is NDCG@10.}
    \Description{Figure shows three plots each comparing box plots for weights of BT without ties, with ties and Plackett-Luce. Top plots show weights for all datasets, the rankings are very similar and the top-performing models are clearly clustered (Seq-KNN, LightGCN, SASRec). The picture is similar for seqential datasets in middle row, but the top-performing algorithms are SASRec and GASATF. In the bottom row (for datasets with long user history) the weights are more smooth and the rankings highly vary across models.}
\end{figure}

We evaluated three Bradley-Terry estimators from Section \ref{sec:bt_estimation}: the classic Zermelo algorithm (MLE), the Bayesian variant, and the Rank Centrality method. On our benchmark data, all three converge to identical weights and rankings. This convergence occurs because our win matrix $W$ is complete and all the models are aimed at maximizing similar loglikelihood. We further use the Bayesian Bradley-Terry model approach (see \eqref{eq:bayes-bt}) due to its principal advantage~-- the availability of confidence intervals for weights, which allows for uncertainty-aware comparisons. In contrast, the Plackett-Luce model often produces different weights and rankings. Figure \ref{fig:btl_comparison} visualizes this divergence, comparing the Bayesian BT weights (with and without tie handling) with PL weights.


 We observe that when the ties are taken into account the weights of consecutive algorithms in the ranking become closer and the quantile boxes overlap more, showing more uncertainty. However, the rankings of BT with and without ties are very similar, indicating that taking ties into account 
 has minimal impact on our benchmark. Moreover, BT is  highly concordant with PL for all and sequential datasets (subfigures (a) and (b)). Both methods clearly identify the same cluster of top-performing algorithms (SASRec, GASATF, etc.). Minor rank variations within this cluster fall within the statistical uncertainty of the estimates. The models disagree more noticeably on datasets with long user histories, where the weights are more smooth. For example, PL moves SASRec from 4th to 6th place and Item-KNN from 9th to 5th. This suggests that while both approaches are valid, when there is more uncertainty in the data, they can give different interpretations of algorithmic performance.

\subsection{Ranking on different dataset classes \label{sec:ranking_classes}}

\begin{table*}[t]
\centering
\caption{NDCG@10-based BT rankings across opposing dataset characteristics. Significant changes (magnitude of at least 3) are highlighted: improvements in \hlgreen{green $\uparrow$} and drops in \hlred{red $\downarrow$}.}
\Description{A comparison table showing the ranking of 14 recommendation algorithms across different dataset characteristics using the NDCG@10 metric. The table contrasts dense vs. sparse datasets, user-heavy vs. item-heavy ratios, long vs. short interaction history, and sequential vs. non-sequential data splits. Significant rank changes are highlighted with colors and arrows. For example, SASRec drops significantly in rank when moving from sequential to non-sequential data, while ALS improves. The table demonstrates how algorithm performance varies depending on specific data properties.}
\label{tab:contrast_pairs_ndcgat10}
\setlength{\tabcolsep}{3pt}
\resizebox{\textwidth}{!}{
\begin{tabular}{cllllllll}
\toprule
 & \multicolumn{2}{c}{Density} & \multicolumn{2}{c}{User-Item Ratio} & \multicolumn{2}{c}{Mean Interaction per User} & \multicolumn{2}{c}{Sequentiality} \\
\cmidrule(lr){2-3} \cmidrule(lr){4-5} \cmidrule(lr){6-7} \cmidrule(lr){8-9}
Rank & \multicolumn{1}{c}{Dense} & \multicolumn{1}{c}{Sparse} & \multicolumn{1}{c}{Large} & \multicolumn{1}{c}{Small} & \multicolumn{1}{c}{Long history} & \multicolumn{1}{c}{Short history} & \multicolumn{1}{c}{Sequential} & \multicolumn{1}{c}{Non-Sequential} \\
\midrule
1 & Seq-KNN & Seq-KNN & LightGCN & LightGCN & Seq-KNN & LightGCN & SASRec & LightGCN \\
2 & LightGCN & LightGCN & SASRec & Seq-KNN & GASATF & Seq-KNN & GASATF & \hlgreen{ALS $\uparrow$} \\
3 & ALS & \hlgreen{GASATF $\uparrow$} & Seq-KNN & GASATF & LightGCN & SASRec & LightGCN & Seq-KNN \\
4 & UltraGCN & SASRec & GASATF & SASRec & SASRec & EASEr & Seq-KNN & EASEr \\
5 & EASEr & UltraGCN & EASEr & EASEr & UltraGCN & \hlred{GASATF $\downarrow$} & EASEr & \hlgreen{UltraGCN $\uparrow$} \\
6 & SASRec & \hlgreen{User-KNN $\uparrow$} & Item-KNN & \hlgreen{UltraGCN $\uparrow$} & EASEr & ALS & PureSVD & Item-KNN \\
7 & SGD MF & EASEr & PureSVD & Item-KNN & ALS & \hlgreen{BPR $\uparrow$} & Item-KNN & \hlgreen{SGD MF $\uparrow$} \\
8 & Item-KNN & Item-KNN & BPR & \hlgreen{ALS $\uparrow$} & PureSVD & \hlred{UltraGCN $\downarrow$} & BPR & BPR \\
9 & GASATF & BPR & UltraGCN & PureSVD & Item-KNN & \hlgreen{User-KNN $\uparrow$} & UltraGCN & \hlgreen{User-KNN $\uparrow$} \\
10 & BPR & PureSVD & SGD MF & SGD MF & SGD MF & Item-KNN & SGD MF & \hlred{SASRec $\downarrow$} \\
11 & PureSVD & \hlred{ALS $\downarrow$} & ALS & \hlred{BPR $\downarrow$} & BPR & SGD MF & ALS & \hlred{GASATF $\downarrow$} \\
12 & User-KNN & \hlred{SGD MF $\downarrow$} & User-KNN & User-KNN & User-KNN & \hlred{PureSVD $\downarrow$} & User-KNN & \hlred{PureSVD $\downarrow$} \\
13 & PopRandom & PopRandom & PopRandom & PopRandom & PopRandom & PopRandom & PopRandom & PopRandom \\
14 & Random & Random & Random & Random & Random & Random & Random & Random \\
\bottomrule
\end{tabular}
}
\end{table*}

\begin{table}[t]
\centering
\caption{Comparison of rankings across all datasets based on NDCG@10. Superscripts indicate rank deviation from BT.}

\Description{A table comparing the rankings of 14 recommendation algorithms across datasets using the NDCG@10 metric. The table lists algorithms sorted by their Bradley-Terry ranking, compared against Sum of Ranks, Mean Rank, and Plackett-Luce model rankings. The top-performing algorithms are Seq-KNN (Rank 1) and LightGCN (Rank 2), showing consistent top-tier performance across all ranking methods (ranks 1-3). Mid-tier algorithms show some variation. Specifically, EASEr is ranked 5th by BT but 1st by Mean Rank, while BPR is ranked 6th by BT but drops to 10th in Sum of Ranks. The lowest-performing algorithms are consistently User-KNN (12), PopRandom (13), and Random (14) across all metrics. Superscripts in the table indicate the numeric deviation of other ranking methods from the BT baseline.}

\label{tab:rank_comparison_all}
\setlength{\tabcolsep}{8pt}

\newcommand{\diff}[1]{\textsuperscript{#1}}
\newcommand{\imp}[2]{\hlgreen{#1\textsuperscript{#2}}}
\newcommand{\drop}[2]{\hlred{#1\textsuperscript{#2}}}

\resizebox{\columnwidth}{!}{
\begin{tabular}{lcccc}
\toprule
Algorithm & \textbf{BT} & Sum & Mean & PL \\ 
\midrule
Seq-KNN & \textbf{1} & 1 & 2\diff{+1} & 1 \\
LightGCN & \textbf{2} & 2 & 3\diff{+1} & 2 \\
SASRec & \textbf{3} & 3 & 4\diff{+1} & 4\diff{+1} \\
GASATF & \textbf{4} & 5\diff{+1} & 5\diff{+1} & 3\diff{-1} \\
EASEr & \textbf{5} & 4\diff{-1} & \imp{1}{-4} & 5 \\
BPR & \textbf{6} & \drop{10}{+4} & 8\diff{+2} & 6 \\
Item-KNN & \textbf{7} & 6\diff{-1} & 6\diff{-1} & 8\diff{+1} \\
UltraGCN & \textbf{8} & 7\diff{-1} & 7\diff{-1} & 7\diff{-1} \\
PureSVD & \textbf{9} & 11\diff{+2} & 11\diff{+2} & 10\diff{+1} \\
ALS & \textbf{10} & 9\diff{-1} & 10 & 11\diff{+1} \\
SGD MF & \textbf{11} & \imp{8}{-3} & 9\diff{-2} & 9\diff{-2} \\
User-KNN & \textbf{12} & 12 & 12 & 12 \\
PopRandom & \textbf{13} & 13 & 13 & 13 \\
Random & \textbf{14} & 14 & 14 & 14 \\
\bottomrule
\end{tabular}
}
\end{table}

\begin{table}[t]
\centering
\caption{Rankings on non-sequential datasets based on NDCG@10. Superscripts indicate rank deviation from BT.}

\Description{A table comparing algorithm rankings on non-sequential datasets based on NDCG@10. Algorithms are sorted by their Bradley-Terry rank. LightGCN is the top performer (Rank 1), followed by ALS (Rank 2) and Seq-KNN (Rank 3). There are notable ranking discrepancies. Specifically, EASEr is ranked 4th by BT but is ranked 1st by both Sum of Ranks and Mean Rank, while UltraGCN is ranked 5th by BT but 2nd by the Plackett-Luce model. The lowest-performing algorithms are consistently PopRandom (13) and Random (14). Superscripts indicate the numeric deviation of the alternative ranking methods (Sum, Mean, PL) from the BT baseline.}

\label{tab:rank_comparison_orderless}
\setlength{\tabcolsep}{8pt}

\newcommand{\diff}[1]{\textsuperscript{#1}}
\newcommand{\imp}[2]{\hlgreen{#1\textsuperscript{#2}}}
\newcommand{\drop}[2]{\hlred{#1\textsuperscript{#2}}}

\resizebox{\columnwidth}{!}{
\begin{tabular}{lcccc}
\toprule
Algorithm & \textbf{BT} & Sum & Mean & PL \\ 
\midrule
LightGCN & \textbf{1} & 2\diff{+1} & 2\diff{+1} & 1 \\
ALS & \textbf{2} & 4\diff{+2} & 4\diff{+2} & 3\diff{+1} \\
Seq-KNN & \textbf{3} & 5\diff{+2} & 5\diff{+2} & 4\diff{+1} \\
EASEr & \textbf{4} & \imp{1}{-3} & \imp{1}{-3} & 5\diff{+1} \\
UltraGCN & \textbf{5} & 3\diff{-2} & 3\diff{-2} & \imp{2}{-3} \\
Item-KNN & \textbf{6} & 7\diff{+1} & 7\diff{+1} & 8\diff{+2} \\
SGD MF & \textbf{7} & 6\diff{-1} & 6\diff{-1} & 6\diff{-1} \\
BPR & \textbf{8} & 9\diff{+1} & 9\diff{+1} & 7\diff{-1} \\
User-KNN & \textbf{9} & 8\diff{-1} & 8\diff{-1} & 9 \\
SASRec & \textbf{10} & 11\diff{+1} & 11\diff{+1} & 11\diff{+1} \\
GASATF & \textbf{11} & 12\diff{+1} & 12\diff{+1} & 10\diff{-1} \\
PureSVD & \textbf{12} & 10\diff{-2} & 10\diff{-2} & 12 \\
PopRandom & \textbf{13} & 13 & 13 & 13 \\
Random & \textbf{14} & 14 & 14 & 14 \\
\bottomrule
\end{tabular}
}
\end{table}

\begin{figure*}[t]
\centering
\begin{subfigure}[t]{0.29\textwidth}
    \includegraphics[height=\heatmapheight]{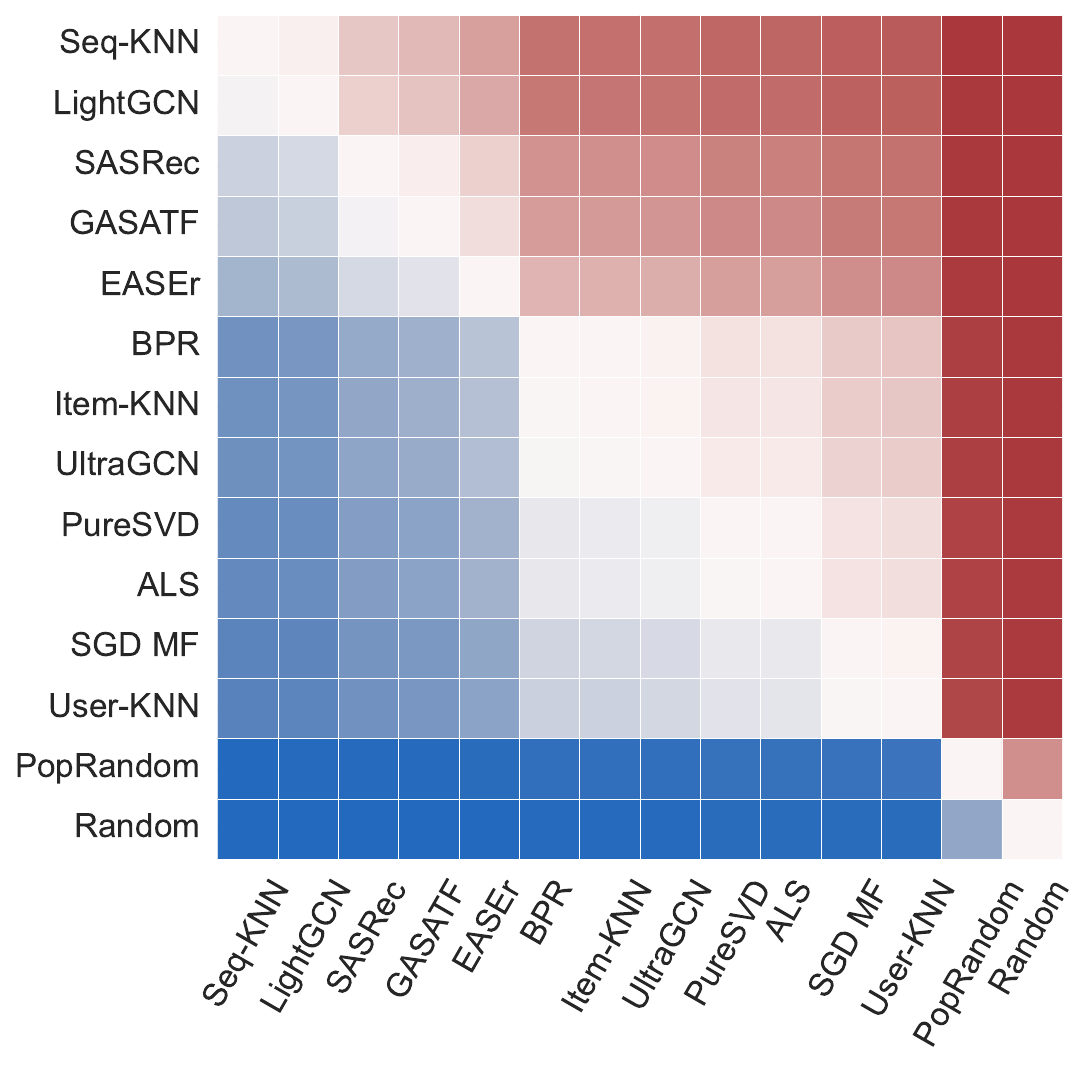}
    \caption{All}
    \label{fig:btl_heatmap_all}
\end{subfigure}
\hspace{0.5em}
\begin{subfigure}[t]{0.29\textwidth}
    \includegraphics[height=\heatmapheight]{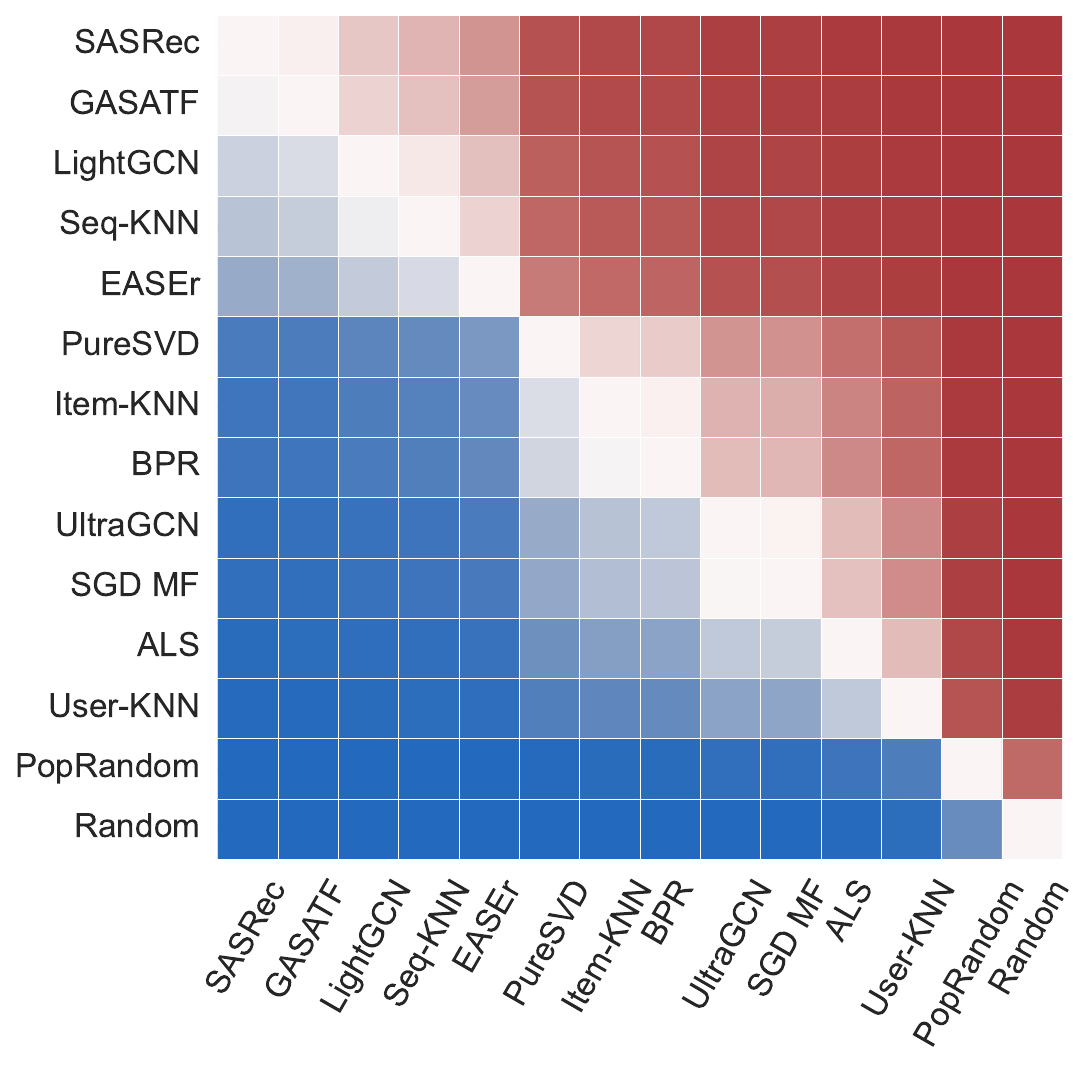}
    \caption{Sequential}
    \label{fig:btl_heatmap_seq}
\end{subfigure}
\hspace{0.5em}
\begin{subfigure}[t]{0.29\textwidth}
    \includegraphics[height=\heatmapheight]{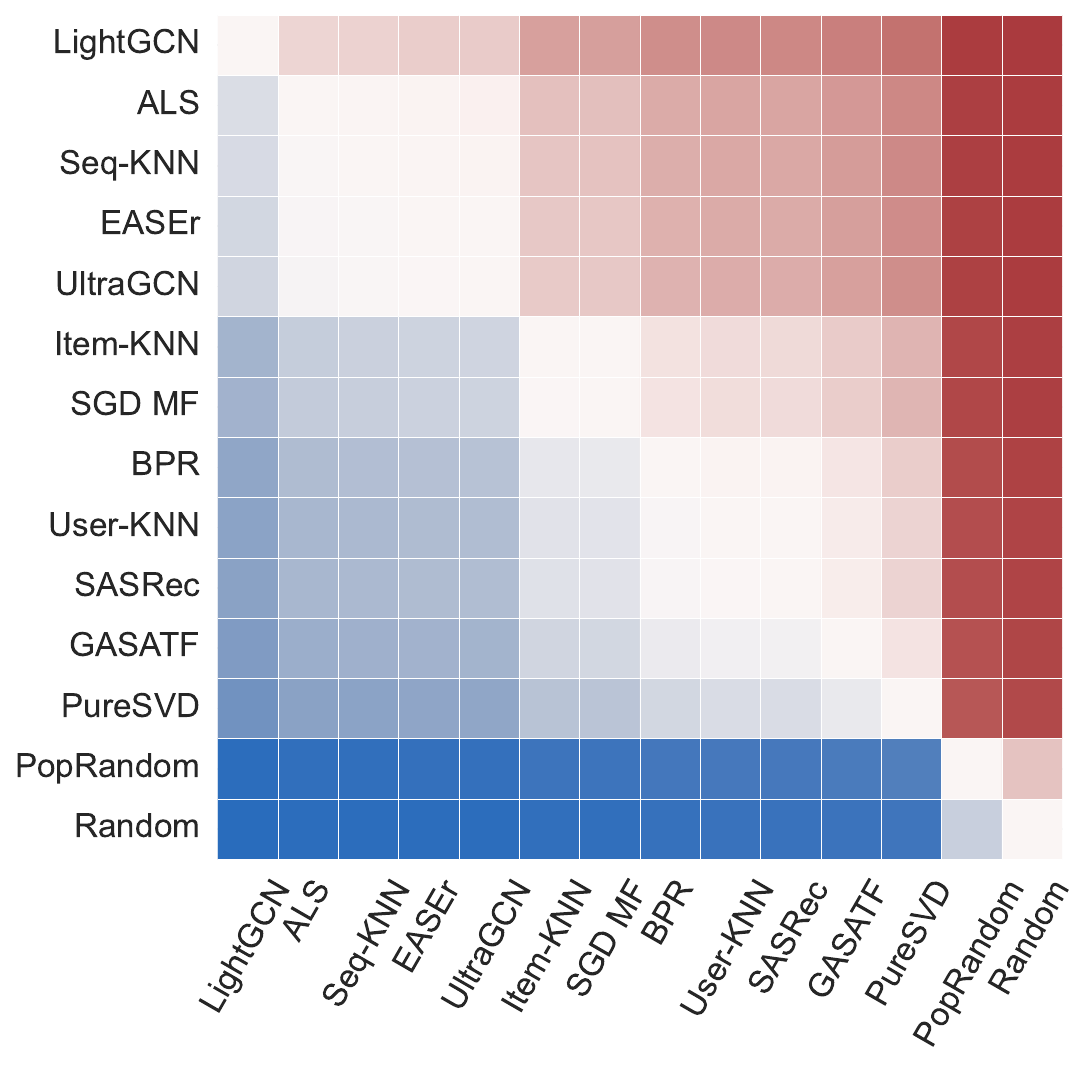}
    \caption{Non-sequential}
    \label{fig:btl_heatmap_non-seq}
\end{subfigure}
\hspace{0.5em}
\begin{subfigure}[t]{0.04\textwidth}
    \includegraphics[height=\heatmapheight]{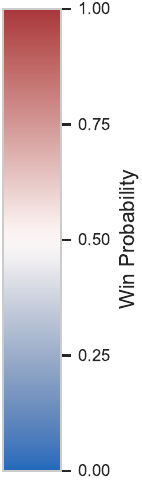}
\end{subfigure}

\caption{Pairwise performance comparison of recommender models. Each cell $(i,j)$ shows the probability that the model in row $i$ outperforms model in column $j$ across all datasets, sequential datasets, and non-sequential datasets.}
\Description{Three heatmaps showing All, Sequential, and Orderless datasets with a shared vertical color scale.}

\Description{Three heatmaps labeled (a) All, (b) Sequential, and (c) Non-sequential, displaying the pairwise win probabilities between 14 recommender algorithms.
The axes represent the algorithms, and the color scale ranges from blue (0.0 probability) to red (1.0 probability).
The rows are ordered by model strength, creating a gradient pattern: the top rows (representing strong models like LightGCN and Seq-KNN) are predominantly red, indicating they consistently outperform other models.
Conversely, the bottom rows (representing weak models like Random and PopRandom) are predominantly blue, indicating they are outperformed by most other models.
The diagonal represents a 50\% probability (self-comparison).}

\label{fig:btl_heatmaps}
\end{figure*}

We analyze how various dataset characteristics influence the performance of recommender systems by using the Bradley-Terry model. Table~\ref{tab:contrast_pairs_ndcgat10} shows that model rankings are not static and change significantly depending on the nature of the data. 
We observe that no single algorithm is the best across all possible scenarios. The most dramatic changes occur in the Sequentiality category. For example, SASRec and GASATF are the clear leaders on sequential datasets, where they hold the first and second ranks. However, their performance drops sharply on non-sequential datasets, where they fall to the 10th and 11th positions. In these non-sequential environments, traditional models like ALS and UltraGCN show significant improvements, with ALS rising from the 11th rank to become the second-best model.

The amount of available user history also plays a critical role in determining which model performs best. On datasets with long history, Seq-KNN and GASATF take the top spots. However, when the interaction history is short, LightGCN becomes the most effective model. In this short history scenario, GASATF drops to 5th place and UltraGCN drops to 8th. This suggests that while some models are very powerful when given a lot of data, they are less robust when information is limited. On sparse datasets, GASATF and User-KNN show improved relative performance compared to their rankings on dense datasets. Furthermore, UltraGCN and ALS improve their rankings on datasets with a low user-to-item ratio, suggesting that these methods are better suited to settings where the item catalog is large relative to the user base.

The heatmaps in Figure~\ref{fig:btl_heatmaps} provide a visualization of the performance clusters through pairwise win probabilities. In Figure~\ref{fig:btl_heatmap_seq} a very dark red block in the top-right corner is clearly observable. This represents a cluster of models including SASRec, GASATF, and LightGCN that have an extremely high probability of beating almost any other algorithm in a sequential context. When we move to Figure~\ref{fig:btl_heatmap_non-seq} for non-sequential data, the cluster changes completely. The dark red areas shift toward LightGCN and ALS, while the previous sequential leaders lose their dominance. This visualization confirms that the competition between models is context-dependent.

Finally, we compare the BT rankings with aggregation methods (Sum and Mean) and discuss why a more sophisticated model is necessary. As shown in Tables~\ref{tab:rank_comparison_all} and~\ref{tab:rank_comparison_orderless}, simple aggregation methods often produce different results because they do not account for the strength of the opponents. In Table~\ref{tab:rank_comparison_all}, the Sum method ranks BPR four positions lower than the BT model. In Table~\ref{tab:rank_comparison_orderless}, both the Sum and Mean methods rank EASEr three positions higher. These rank deviations happen because simple methods treat every win in a dataset as equal. The BT model provides a more accurate hierarchy by considering what opponent an algorithm beats and how strong that opponent is. This makes the BT framework a much more reliable tool for understanding the true strengths of different recommendation algorithms across diverse data environments.

\subsection{Separation strength}

We check the separation strength of the ranking by providing an ablation study on sequential datasets. We shuffle timestamps in the datasets and compare the ranking before and after shuffling. Contrary to the setup in~\cite{klenitskiy2024sequential}, we apply timestamp shuffling to the training sequences and retrain the models.

\begin{figure}
    \centering
    \begin{subfigure}[t]{0.47\textwidth}
    \centering
    \includegraphics[width=1\linewidth]{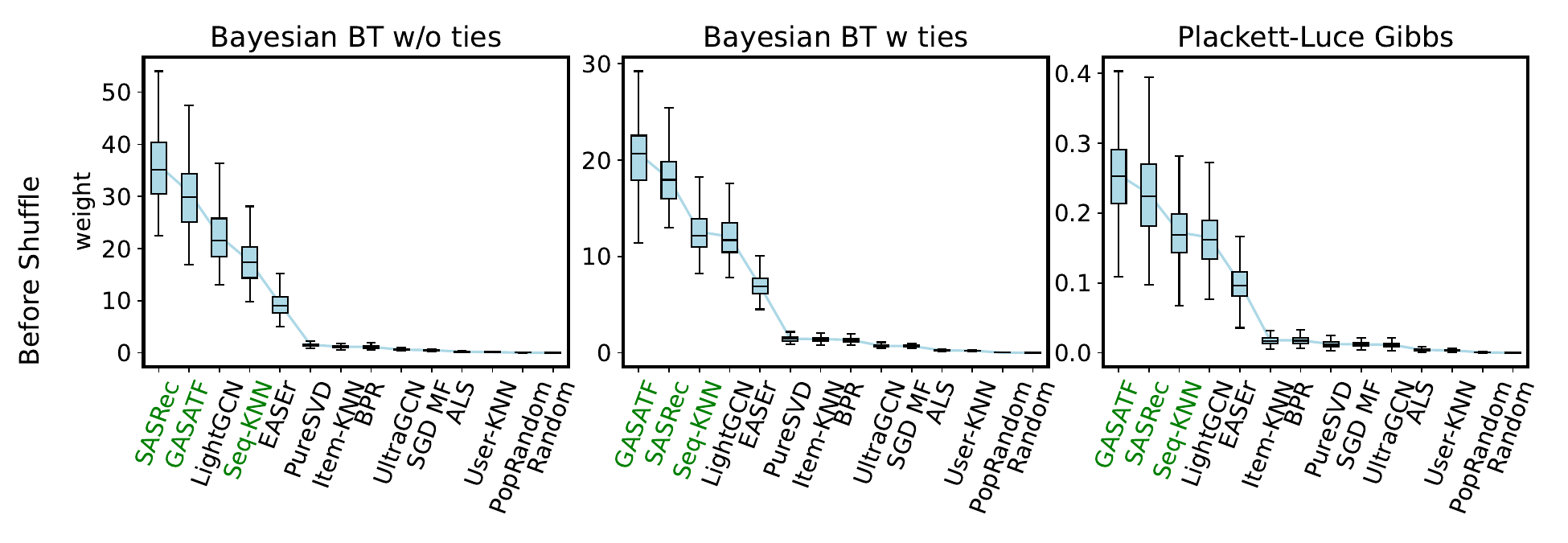}
    \end{subfigure}%
    \vfill
    \begin{subfigure}[t]{0.47\textwidth}
    \centering
    \includegraphics[width=1\linewidth]{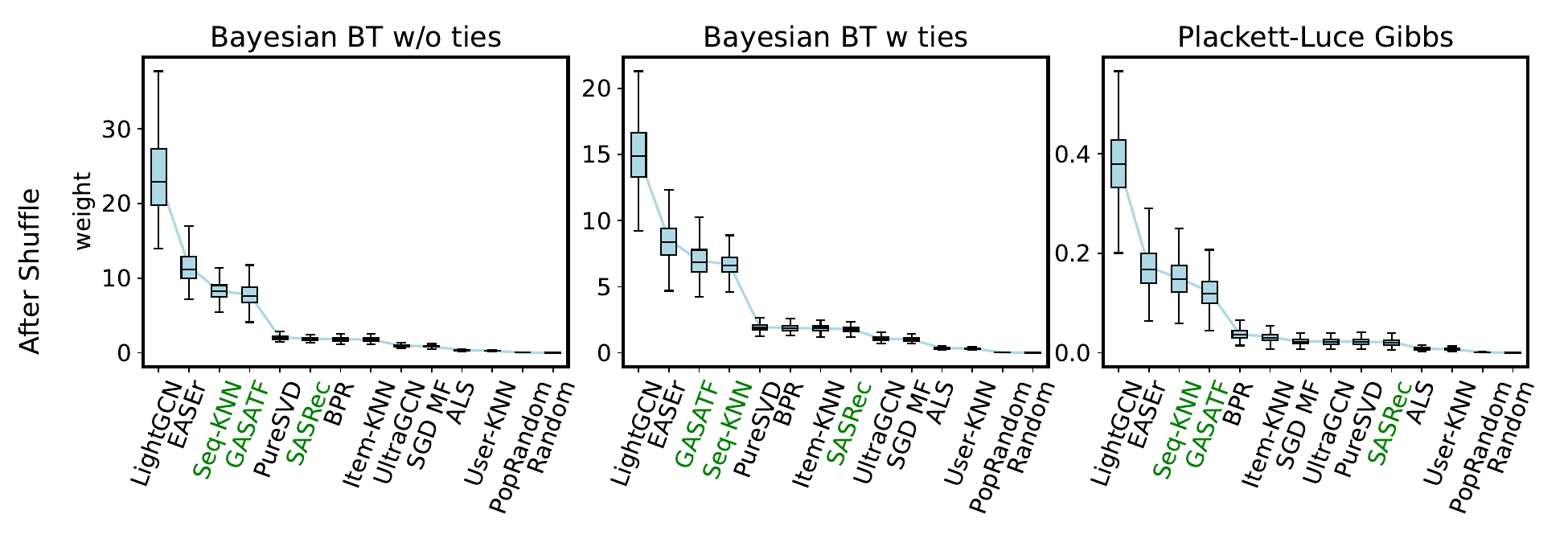}
    \end{subfigure}%
    \caption{\label{fig:separation:seq} Comparison of weights of different models on sequential datasets. The metric for ranking is NDCG@10. Top row - before shuffling timestamps, bottom row - after. Sequential models are highlighted in green.}
    \Description{Comparison of weights of different models on sequential datasets. The metric for ranking is NDCG@10. The top row presents the results before shuffling timestamps, and the bottom row shows the results after shuffling timestamps. In the top row, sequential models are at the top of the ranking, with Seq-KNN consistently being the lowest-ranked among sequential models. In the bottom row, sequential models move to the middle of the ranking, and SASRec becomes the lowest-ranked sequential model.}
\end{figure}

Figure~\ref{fig:separation:seq} presents the results. As expected, sequential models dominate sequential datasets before timestamp shuffling, but lose this advantage and become less separated afterward. The ranking among sequential models also changes: SASRec is among the top-2 sequential models for all rankings before shuffling, but after shuffling, Seq-KNN and GASATF become ranked higher than SASRec. These models are more stable under different setups, while the rank of SASRec substantially degrades after shuffling. The  weights of BT model decrease after shuffling, since the obvious leaders -- sequential models -- lose their advantage from the sequential data structure.

\section{Bradley-Terry model with covariates \label{sec:main_bt_covariates}}

A huge disadvantage of a simple BT model is its inability to model different rankings on different datasets and, hence, predict a specific ranking on a previously unseen dataset (which may drastically differ from the global BT ranking obtained on the training datasets). This can be amended by introducing dataset characteristics as covariates of the pairwise comparisons and exploit a covariate-adjusted BT model. We experiment with different covariate-based approaches (described in Sections~\ref{sec:bt_covariate}, \ref{sec:bt_trees}) in order to predict algorithms ranking on new datasets and evaluate their prediction accuracy in Section~\ref{sec:bt_covariate_eval}, providing the guidelines for strong baseline selection for the practitioners.


\subsection{Covariate-adjusted Bradley-Terry model \label{sec:bt_covariate}}

In the standard BT model $P(i\succ j)$ can be rewritten as $P(i\succ j) = \sigma (\theta_i - \theta_j)$, where $1 \leq i, j \le n$ are players and $\theta_i, \theta_j \in \mathbb{R}$ are their corresponding strengths, $\sigma(z) = \frac{1}{1+e^{-z}}$. Now suppose that for each pairwise comparison exists a covariate vector $x \in \mathbb{R}^d$, influencing players' strengths. Covariate-adjusted BT models (and other pairwise Learning-to-Rank models) typically integrate it via a linear combination, what renders a huge space for various approaches and remains convenient for theoretical analysis and interpretation (see \cite{ schauberger2019btllasso, fan2024uncertainty, casalicchio2015subject}
). 
Namely, player $i$  is now characterized by a strength parameter $\beta_{i0} \in \mathbb{R}$ and covariate weights $\beta_i = (\beta_{i1}, \dots, \beta_{id})^\top \in \mathbb{R}^d$, quantifying different covariates' influence on $i$'s advantage. Thus, we can define
\[
P(i\succ j \,|\,  x) = \sigma (\beta_{i0} - \beta_{j0} + \langle x, \beta_i\rangle - \langle x, \beta_j\rangle).
\]

This model has $\mathcal{B} = \cup_i \{\beta_{i0}, \beta_i\}, |\mathcal{B}| = n(d+1)$ parameters and requires training data $R = \{(i_k, j_k, y_{k}, x_k)\}_k$, where $1 \leq i_k, j_k \le n$ are players, $ y_{k} \in \{0, 1\}$ is a player $i_k$ win indicator and $x_k \in \mathbb{R}^d$ is a covariate vector. It can be fit by maximizing log-likelihood $\ell(\mathcal{B}) = \sum_{k} (y_k \log P(i_k\succ j_k \,|\,  x_k)  + (1 - y_k)\log(1 -  P(i_k\succ j_k \,|\,  x_k))$ via any appropriate optimization method under identifiability constraints $\forall j\sum _{i=1}^n\beta_{ij} = 0$.

However, 
this simple model is prone to overfitting to the global BT ranking with static and clearly separable strengths and little to no covariate effect. This can be amended by introducing L1 penalty following~\cite{schauberger2019btllasso} as $L(\mathcal{B}) = \lambda_0 \sum_{i < j}\sqrt{(\beta_{i0} - \beta_{j0})^2 + \varepsilon} + \lambda_1 \sum_{k=1}^d\sum_{i < j}\sqrt{(\beta_{ik} - \beta_{jk})^2 + \varepsilon}$, where $\varepsilon > 0$, the first term regulates the closeness of overall strengths and the second term~-- the closeness of covariate effects weights. 
The final target function is $\ell(\mathcal{B}) - L(\mathcal{B}) \rightarrow \max$. We
optimize the penalized objective with L-BFGS-B in the subspace $\forall j\sum _{i=1}^n\beta_{ij} = 0$ by assuming $\beta_{nj} = -\sum_{i=1}^{n-1}\beta_{ij}$. In our setup, we consider log-normalized numeric and binary categorical dataset characteristics as covariates and predict ranking on a new dataset with covariates $x$ by sorting algorithms' strengths on this particular dataset, precisely $\beta_{i0} + \langle x, \beta_{i}\rangle$.


\subsection{Bradley-Terry trees \label{sec:bt_trees}}

\begin{figure}[htb]
    \centering
        \includegraphics[width=1\linewidth]{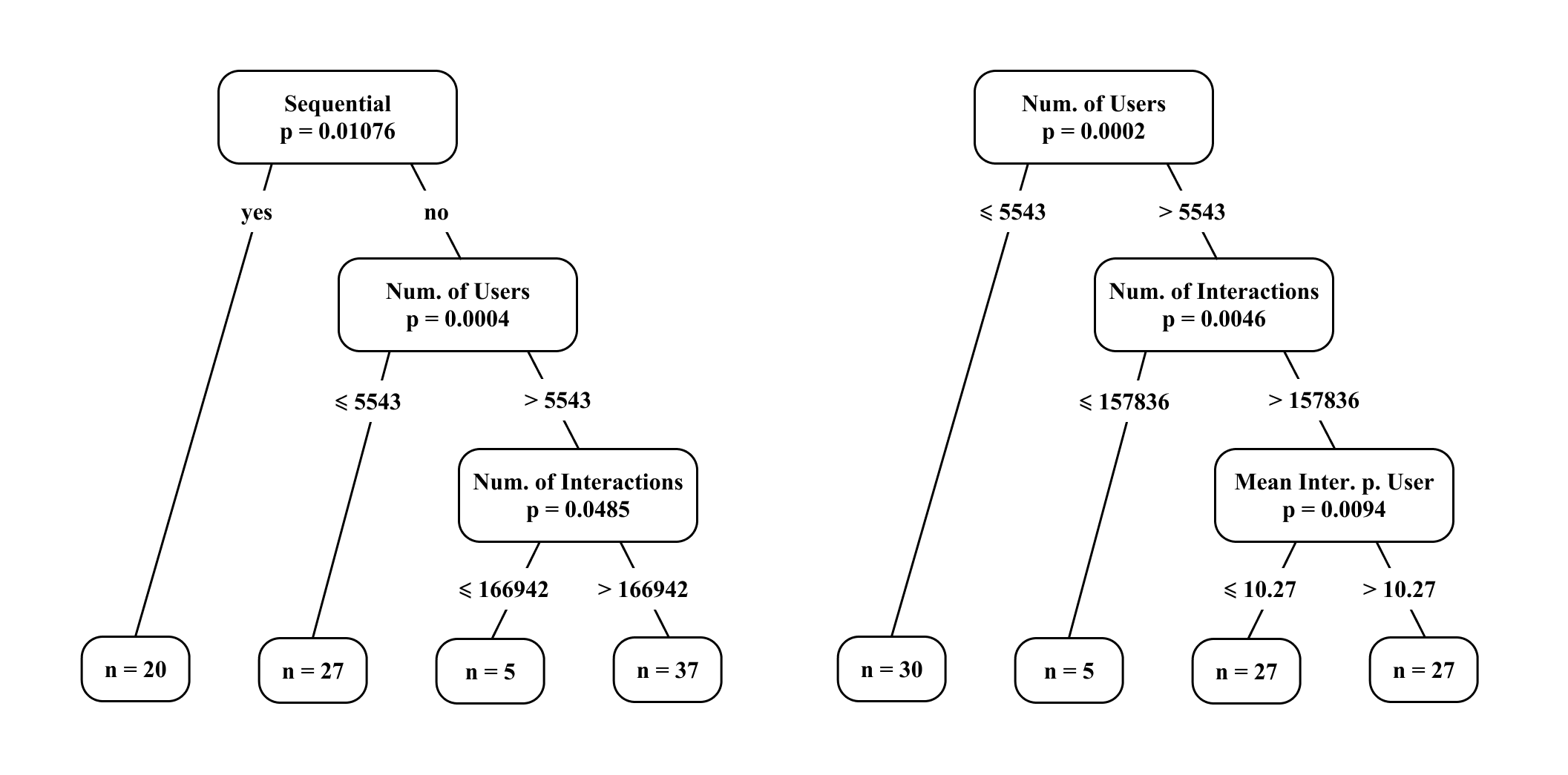}
        \caption{BT trees for datasets. Each node displays the split covariate and its $p$-value, with split conditions on the edges; leaf sample sizes are denoted by $n$.}
        \label{fig:trees}
        \Description{The image shows two BT trees for datasets. The left tree first separates datasets by sequentiality, then non-sequential datasets are separated by number of users and datasets with high number of users are divided by number of interaction per user. The right tree divides by number of users, number of interactions and mean interaction per user. The figure shows p-values for each separation and number of datasets in resulting nodes.}
\end{figure}

A non-parametric niche alternative to the parametric approach presented above is the recursive covariate-dependent partitioning via Bradley-Terry trees. It can identify groups of datasets with distinct algorithm rankings in a data‑driven way \cite{zeileis2008modelpartitioning}, thus yielding interpretable schemes that can predict algorithms rankings based on dataset characteristics. 
In order to build the trees for our data, we use numeric characteristics as numeric covariates and sequentiality as a categorical one, following the algorithm of \cite{zeileis2008modelpartitioning}. We tried two variants of model: first with an enforced root split by sequentiality (acknowledged as an important characteristic, see e.g. \cite{shevchenko2024benchmarking}), and then without any constraints. Figure~\ref{fig:trees} shows the resulting trees: left with forced sequentiality split, right without adjustments. We observe that for sequential datasets rankings remain largely stable, whereas for non‑sequential ones they depend mostly on the number of users and the number of interactions. Finally, we can use the constructed trees to predict ranking on a new dataset by simply traversing through them with its characteristics with zero computational cost.


\subsection{Ranking on unseen dataset \label{sec:bt_covariate_eval}}

In order to evaluate different models' ability to select strong baselines for different previously unobserved datasets, we use the following pipeline. We randomly choose 10 holdout datasets, fit models on the remaining training datasets, and predict rankings for each holdout dataset accordingly. Evaluation metrics are: (i) hit rate of the top‑1 predicted algorithm being among the top‑1/2/3 ground‑truth algorithms on the holdout dataset (denoted as top-1/2/3 hits), and (ii) the overlap of top‑2/3/5 predicted algorithms and the top‑2/3/5 ground‑truth algorithms correspondingly (denoted as top-2/3/5 overlap). Results are averaged over 5 runs.

We compare trivial mean aggregation (obtaining the same ranking for all holdout datasets by averaging the metric on training datasets, denoted as Mean), simple Bradley-Terry model (obtaining the same ranking for all holdout datasets by fitting a BT model on training datasets, denoted as BT), Bradley-Terry trees (as described in Section~\ref{sec:bt_trees}, denoted as BT tree) and covariate-adjusted Bradley-Terry model with fusion regularization (as described in Section~\ref{sec:bt_covariate}, denoted as Cov. BT).

As shown in Table~\ref{tab:prediction_accuracy}, with sufficient training data (Train (79) / Holdout (10)) all of BT variations consistently outperform the trivial Mean aggregation and demonstrate a decent performance overall, predicting top-1 algorithm with 0.24-0.28 accuracy and approximately 3.3 out of top-5 algorithms on average. It is also important to note that with the decreasing amount of data for fitting a simple BT model (i.e., less training and more holdout datasets), its ability to accurately predict top-1 algorithm decreases in comparison with covariate models, while the ability to predict strong baselines remains on par with others. Such results show that covariate-adjusted models are more accurate, especially with fewer comparisons, but for our purposes BT model is a strong and reliable method. Moreover, this highlights the importance of using the large and diverse set of datasets we have gathered. 

\begin{table}[h]
\centering
\caption{\label{tab:prediction_accuracy} Prediction accuracy metrics for holdout sizes of 10 and 30 datasets.}
\begin{tabular}{lcccc}
\toprule
\multicolumn{1}{c}{} & \multicolumn{4}{c}{Model} \\
\cmidrule{2-5}
Metric & Mean & BT & Cov. BT & BT tree \\
\midrule
\multicolumn{5}{c}{\textbf{Train (79) / Holdout (10)}} \\
\midrule
top-1 hits   & 0.02 & 0.24 & 0.28 & 0.26 \\
top-2 hits & 0.12 & 0.58 & 0.62 & 0.50 \\
top-3 hits & 0.16 & 0.78 & 0.78 & 0.64 \\
top-2 overlap  & 0.62 & 0.86 & 0.96 & 0.86 \\
top-3 overlap  & 1.38 & 1.64 & 1.72 & 1.64 \\
top-5 overlap  & 3.32 & 3.32 & 3.28 & 2.88 \\
\midrule
\multicolumn{5}{c}{\textbf{Train (59) / Holdout (30)}} \\
\midrule
top-1 hits   & 0.08 & 0.15 & 0.24 & 0.13 \\
top-2 hits & 0.17 & 0.47 & 0.61 & 0.40 \\
top-3 hits & 0.25 & 0.63 & 0.71 & 0.56 \\
top-2 overlap & 0.63 & 0.87 & 0.97 & 0.77 \\
top-3 overlap  & 1.41 & 1.64 & 1.64 & 1.42 \\
top-5 overlap  & 3.40 & 3.40 & 3.28 & 3.00 \\
\bottomrule
\end{tabular}
\end{table}

Finally, we observe the following phenomenon: although covariate-adjusted Bradley-Terry models provide an overall more accurate ranking, simple Bradley-Terry ranking (providing the same top-5 algorithms for all new datasets) is actually sufficient for strong baseline prediction. 
To show this, we ran our pipeline calculating average Kendall correlation between rankings obtained by different BT models and ground-truth ranking by the metic value on the dataset as well as MAP@5, NDCG@5 and top-5 overlap (as metrics reflecting specifically the ability to obtain strong baselines). The results presented in Table~\ref{tab:ranking_accuracy} confirm our hypothesis.

\begin{table}[h]
\centering
\caption{\label{tab:ranking_accuracy} Ranking accuracy metrics.}
\begin{tabular}{lcccc}
\toprule
Metric & Mean & BT & Cov. BT & BT tree \\
\midrule
Kendall's $\tau$  & 0.438 & 0.489 & \textbf{0.572} & 0.501 \\
MAP@5    & 0.689 & 0.872 & \textbf{0.873} & 0.869 \\
NDCG@5   & 0.575 & 0.709 & \textbf{0.732} & 0.693 \\
top-5 overlap    & 3.10  & 3.32  & \textbf{3.42}  & 3.18 \\
\bottomrule
\end{tabular}
\end{table}


Overall, these results show that in order to obtain an adequate set of strong baselines for a new dataset, the global BT ranking from Table~\ref{tab:rank_comparison_all} can be used. However, if a more accurate ranking prediction is needed, the covariate-augmented BT model is preferable.

\section{Discussion}
A major limitation of the Bradley-Terry model is that it compares algorithms based on the number of wins, but does not take the possibility of insignificant difference between algorithm performance (i.e. a ``tie'') into account. This can be done via simply introducing the parameter regulating the probability of a tie (see \cite{davidson1970extending}), as well as through more complex methods taking into account the magnitude of the strengths (see \cite{glickman2025paired}). The considerate shortcomings of pointwise estimates have also been addressed in the literature \cite{haldar2025ltr} with a true-pairwise learning-to-rank algorithm, training a bivariate MLP and deploying BT on pairwise scores. 

Our work has a significant potential in a variety of applications in both research and practice. We provide an open repository with 14 RecSys algorithms implementations and ready-to-use framework for hyperparameter optimization, evaluation on 89 preprocessed datasets, results aggregation and analysis via Bradley-Terry model. 
Our work can also serve as a practical guide for selecting strong algorithms for a given dataset based on its features.

\section{Conclusion}


The paper proposes a methodology for ranking recommendation algorithms based on the Bradley--Terry model. The method is supplemented by the transitive triplets metric for assessing ranking consistency and robustness to missing data. Experiments reveal that rankings vary significantly with dataset characteristics such as sequentiality and sparsity. Building on this observation, BT trees and covariate-adjusted BT enable dataset-aware baseline selection by predicting algorithm rankings for a target dataset from its characteristics, without running the candidate models. Our approach provides a robust and interpretable comparison that is substantially more stable than simple metric aggregation methods, even in the presence of missing data. The resulting open benchmark and analysis tools provide a practical foundation for reproducible and more reliable algorithm comparison in recommender systems.



\begin{acks}
The work was supported by the grant for research centers in the field of AI provided by the Ministry of Economic Development of the Russian Federation in accordance with the agreement \\ 000000C313925P4E0002 and the agreement with HSE University № 139-15-2025-009. This research was supported in part through computational resources of HPC facilities at HSE University \cite{kostenetskiy2021hpc}. 
\end{acks}

\bibliographystyle{ACM-Reference-Format}
\bibliography{sample-base}

\appendix
\onecolumn
\section{Supplementary tables}

\begin{table}[h]
  \caption{Recommendation algorithms used in the study}
  \label{table:rec_algorithms}
  \begin{tabularx}{\linewidth}{lXl}
    \toprule
    Algorithm & Description & Implementation\\
    \midrule
    Random & Recommends items by sampling uniformly at random & self implemented\\
    PopRandom & Recommends items by sampling in proportion to their frequency & self implemented\\
    User-KNN & Recommends items from similar users & self implemented\\
    Item-KNN & Recommends items similar to the user’s history & self implemented\\
    Seq-KNN & Recommends items from similar sequential user histories & self implemented\\
    SGD MF & One batch SGD with adaptive negative sampling \cite{randle_negative_sampling} and folding-in procedure & self implemented\\
    BPR & Based on \cite{rendle2009bpr} with uniform negative sampling and folding-in procedure & self-implemented \\
    ALS & Based on \cite{hu2008collaborative, takacs2011applications} & ``Implicit'' package\\
    PureSVD & Based on~\cite{puresvd}, using SVD routine from~\cite{tsyganov2026fastersvdvia} & self implemented\\
    EASEr & Based on \cite{steck2019easer} & self-implemented \\
    GASATF & Based on \cite{gasatf} & self implemented\\
    LightGCN & Original LightGCN~\cite{lightgcn} model with BPR objective & Pytorch-geometric \\
    UltraGCN & Original UltraGCN~\cite{mao2021ultragcn} model with folding-in procedure. We also tried regularization as in~\cite{tsyganov2026matrix} & self implemented.\\
    SASRec & SASRec~\cite{kang2018sasrec} with Scalable Cross-Entropy~\cite{mezentsev2024sce} and alpha-beta reparametrization of bucket sizes & from official repository \\
    \bottomrule
\end{tabularx}
\end{table}

\begin{table}[H]
\centering
\caption{Kendall's $\tau$ between BT rankings and baseline aggregation methods. The highest correlation for each group is \textbf{bold}.}
\label{tab:kendall_corr}
\Description{A table displaying Kendall rank correlation coefficients between the Bradley-Terry ranking model and three baseline aggregation methods: Mean, Sum, and Plackett-Luce. The correlations are computed for NDCG@10, HitRate@10, and Coverage@10. All metrics show strong positive correlations (mostly > 0.8), indicating that the BT model effectively captures the ranking order of standard metrics. Plackett-Luce (PL) often achieves the highest correlation, particularly for sparse datasets and coverage metrics.}
\begin{tabular}{lccccccccc}
\toprule
 & \multicolumn{3}{c}{NDCG@10} & \multicolumn{3}{c}{HitRate@10} & \multicolumn{3}{c}{Coverage@10} \\
\cmidrule(lr){2-4} \cmidrule(lr){5-7} \cmidrule(lr){8-10}
Dataset group & Mean & Sum & PL & Mean & Sum & PL & Mean & Sum & PL \\
\midrule
All & 0.802 & 0.824 & \textbf{0.912} & 0.824 & 0.846 & \textbf{0.868} & 0.934 & 0.890 & \textbf{0.956} \\
Sequential & 0.890 & \textbf{0.912} & 0.890 & 0.846 & 0.868 & \textbf{0.912} & 0.912 & 0.934 & \textbf{0.978} \\
Non-sequential & 0.802 & 0.802 & \textbf{0.868} & 0.868 & 0.868 & \textbf{0.912} & 0.846 & 0.846 & \textbf{0.868} \\
Dense & \textbf{0.824} & \textbf{0.824} & 0.802 & \textbf{0.824} & \textbf{0.824} & \textbf{0.824} & 0.846 & 0.846 & \textbf{0.890} \\
Sparse & 0.692 & 0.736 & \textbf{0.912} & 0.736 & 0.758 & \textbf{0.956} & \textbf{0.978} & 0.758 & 0.956 \\
Mostly users & 0.692 & 0.714 & \textbf{0.912} & 0.692 & 0.692 & \textbf{0.912} & \textbf{0.934} & 0.846 & 0.912 \\
Mostly items & 0.780 & \textbf{0.824} & 0.758 & 0.758 & 0.802 & \textbf{0.868} & \textbf{0.890} & 0.846 & \textbf{0.890} \\
Long history & 0.736 & \textbf{0.780} & 0.758 & 0.758 & 0.802 & 0.758 & 0.802 & 0.824 & \textbf{0.934} \\
Short history & 0.626 & 0.648 & \textbf{0.890} & 0.648 & 0.670 & \textbf{0.846} & 0.912 & 0.890 & \textbf{0.956} \\
\bottomrule
\end{tabular}
\end{table}

\begin{table}
\centering
\caption{Recall@10-based BT rankings across opposing datasets. Significant improvements are \hlgreen{green $\uparrow$}, drops are \hlred{red $\downarrow$}.}
\label{tab:contrast_pairs_recall10}
\resizebox{0.8\textwidth}{!}{
\begin{tabular}{cllllllll}
\toprule
 & \multicolumn{2}{c}{Density} & \multicolumn{2}{c}{User-Item Ratio} & \multicolumn{2}{c}{Mean Interaction per User} & \multicolumn{2}{c}{Sequentiality} \\
\cmidrule(lr){2-3} \cmidrule(lr){4-5} \cmidrule(lr){6-7} \cmidrule(lr){8-9}
Rank & \multicolumn{1}{c}{Dense} & \multicolumn{1}{c}{Sparse} & \multicolumn{1}{c}{Large} & \multicolumn{1}{c}{Small} & \multicolumn{1}{c}{Long history} & \multicolumn{1}{c}{Short history} & \multicolumn{1}{c}{Sequential} & \multicolumn{1}{c}{Non-Sequential} \\
\midrule
1 & Seq-KNN & Seq-KNN & LightGCN & LightGCN & Seq-KNN & Seq-KNN & GASATF & LightGCN \\
2 & LightGCN & LightGCN & Seq-KNN & GASATF & GASATF & LightGCN & SASRec & \hlgreen{ALS $\uparrow$} \\
3 & ALS & \hlgreen{SASRec $\uparrow$} & SASRec & Seq-KNN & LightGCN & SASRec & LightGCN & \hlgreen{UltraGCN $\uparrow$} \\
4 & UltraGCN & \hlgreen{GASATF $\uparrow$} & GASATF & SASRec & SASRec & GASATF & Seq-KNN & Easer \\
5 & Easer & Easer & Easer & \hlgreen{UltraGCN $\uparrow$} & UltraGCN & Easer & Easer & Seq-KNN \\
6 & SGD MF & UltraGCN & Item-KNN & Easer & ALS & \hlgreen{BPR $\uparrow$} & Pure SVD & \hlgreen{SGD MF $\uparrow$} \\
7 & SASRec & BPR & BPR & Item-KNN & Easer & ALS & BPR & Item-KNN \\
8 & GASATF & Item-KNN & Pure SVD & ALS & SGD MF & UltraGCN & Item-KNN & BPR \\
9 & Item-KNN & User-KNN & UltraGCN & SGD MF & Item-KNN & User-KNN & SGD MF & \hlred{SASRec $\downarrow$}\\
10 & BPR & Pure SVD & SGD MF & \hlred{BPR $\downarrow$} & Pure SVD & SGD MF & UltraGCN & User-KNN \\
11 & User-KNN & \hlred{ALS $\downarrow$} & ALS & Pure SVD & BPR & Item-KNN & ALS & \hlred{GASATF $\downarrow$}\\
12 & Pure SVD & \hlred{SGD MF $\downarrow$} & User-KNN & User-KNN & User-KNN & Pure SVD & User-KNN & \hlred{Pure SVD $\downarrow$} \\
13 & PopRandom & PopRandom & PopRandom & PopRandom & PopRandom & PopRandom & PopRandom & PopRandom \\
14 & Random & Random & Random & Random & Random & Random & Random & Random \\
\bottomrule
\end{tabular}
}
\end{table}


\begin{table}
  \centering
  \caption{Left: statistics of the datasets used in the study. Right: number of missing trials for the specified algorithm.}
  \label{table:dataset_stats_clean}
  \tiny 
  \setlength{\tabcolsep}{2pt}
  \begin{tabular}{l|rrrrrrr|rrrrrrr}
    \toprule
    \thead{Dataset} & \thead{Number \\ of Users} & \thead{Number \\of Items} & \thead{Number of \\Interactions} & \thead{User-Item \\ Ratio} & \thead{Density\\ ($\%$)} & \thead{Mean \\Interaction \\ per User} & \thead{Mean \\Interaction\\per Item} & \thead{BPR} & \thead{EASEr} & \thead{GASATF} &  \thead{LightGCN} & \thead{SGD MF} & \thead{SASRec} & \thead{UltraGCN}\\
    \midrule
Amazon2014-Instant-Video & 5130 & 1685 & 37126 & 3.04 & 0.43 & 7.24 & 22.03  & - & - & - & - & - & - & -  \\
Amazon2014-Apps-For-Android & 87271 & 13209 & 752937 & 6.61 & 0.07 & 8.63 & 57.0  & - & - & - & - & - & - & -  \\
Amazon2014-Automotive & 2928 & 1835 & 20473 & 1.6 & 0.38 & 6.99 & 11.16  & - & - & - & - & - & - & -  \\
Amazon2014-Baby & 19445 & 7050 & 160792 & 2.76 & 0.12 & 8.27 & 22.81  & - & - & - & - & - & - & -  \\
Amazon2014-Beauty & 22363 & 12101 & 198502 & 1.85 & 0.07 & 8.88 & 16.4  & - & - & - & - & - & - & -  \\
Amazon2014-Books & 603668 & 367982 & 8898041 & 1.64 & 0.00 & 14.74 & 24.18  & 127 & - & 147 & 192 & 165 & 200 & 190  \\
Amazon2014-CDs-Vinyl & 75258 & 64443 & 1097592 & 1.17 & 0.02 & 14.58 & 17.03  & - & - & - & - & - & - & -  \\
Amazon2014-Cell-Phones-And-Accessories & 27879 & 10429 & 194439 & 2.67 & 0.07 & 6.97 & 18.64  & - & - & - & - & - & - & -  \\
Amazon2014-Clothing-Shoes-And-Jewelry & 39387 & 23033 & 278677 & 1.71 & 0.03 & 7.08 & 12.1  & - & - & - & - & - & 200 & -  \\
Amazon2014-Digital-Music & 5541 & 3568 & 64706 & 1.55 & 0.33 & 11.68 & 18.14  & - & - & - & - & - & - & -  \\
Amazon2014-Electronics & 192403 & 63001 & 1689188 & 3.05 & 0.01 & 8.78 & 26.81  & - & 190 & - & - & - & - & -  \\
Amazon2014-Grocery-And-Gourmet & 14681 & 8713 & 151254 & 1.68 & 0.12 & 10.3 & 17.36  & - & - & - & - & - & - & -  \\
Amazon2014-Health-And-Personal-Care & 38609 & 18534 & 346355 & 2.08 & 0.05 & 8.97 & 18.69  & - & - & - & - & - & - & -  \\
Amazon2014-Home-And-Kitchen & 66519 & 28237 & 551682 & 2.36 & 0.03 & 8.29 & 19.54  & - & 190 & - & - & - & - & -  \\
Amazon2014-Kindle-Store & 68223 & 61934 & 982619 & 1.1 & 0.02 & 14.4 & 15.87  & - & 190 & - & - & - & - & -  \\
Amazon2014-Movies-And-TV & 123960 & 50052 & 1697533 & 2.48 & 0.03 & 13.69 & 33.92  & - & 190 & - & - & - & - & -  \\
Amazon2014-Musical-Instruments & 1429 & 900 & 10261 & 1.59 & 0.8 & 7.18 & 11.4  & - & - & - & - & - & - & -  \\
Amazon2014-Office-Products & 4905 & 2420 & 53258 & 2.03 & 0.45 & 10.86 & 22.01  & - & - & - & - & - & - & -  \\
Amazon2014-Patio-Lawn-And-Garden & 1686 & 962 & 13272 & 1.75 & 0.82 & 7.87 & 13.8  & - & - & - & - & - & - & -  \\
Amazon2014-Pet-Supplies & 19856 & 8510 & 157836 & 2.33 & 0.09 & 7.95 & 18.55  & - & - & - & - & - & - & -  \\
Amazon2014-Sports-And-Outdoors & 35598 & 18357 & 296337 & 1.94 & 0.05 & 8.32 & 16.14  & - & - & - & - & - & - & -  \\
Amazon2014-Tools-And-Home-Improvement & 16638 & 10217 & 134476 & 1.63 & 0.08 & 8.08 & 13.16  & - & - & - & - & - & - & -  \\
Amazon2014-Toys-And-Games & 19412 & 11924 & 167597 & 1.63 & 0.07 & 8.63 & 14.06  & - & - & - & - & - & - & -  \\
Amazon2014-Video-Games & 24303 & 10672 & 231780 & 2.28 & 0.09 & 9.54 & 21.72  & - & - & - & - & - & - & -  \\
Amazon2018-Arts-Crafts-And-Sewing & 55969 & 22611 & 438698 & 2.48 & 0.03 & 7.84 & 19.4  & - & - & - & - & - & - & -  \\
Amazon2018-Automotive & 193328 & 78977 & 1637345 & 2.45 & 0.01 & 8.47 & 20.73  & - & 190 & - & 192 & - & - & -  \\
Amazon2018-CDs-And-Vinyl & 112134 & 73303 & 1399872 & 1.53 & 0.02 & 12.48 & 19.1  & - & 190 & - & - & - & - & -  \\
Amazon2018-Cell-Phones-And-Accessories & 157037 & 47983 & 1119546 & 3.27 & 0.01 & 7.13 & 23.33  & - & 190 & - & - & - & - & -  \\
Amazon2018-Digital-Music & 16252 & 11269 & 142820 & 1.44 & 0.08 & 8.79 & 12.67  & - & - & - & - & - & - & -  \\
Amazon2018-Electronics & 728489 & 159729 & 6526514 & 4.56 & 0.01 & 8.96 & 40.86  & 127 & 190 & 147 & 192 & 165 & 200 & 190  \\
Amazon2018-Gift-Cards & 456 & 147 & 2953 & 3.1 & 4.41 & 6.48 & 20.09  & - & - & - & - & - & - & -  \\
Amazon2018-Grocery-And-Gourmet-Food & 127278 & 40994 & 1063722 & 3.1 & 0.02 & 8.36 & 25.95  & - & 190 & - & - & - & - & -  \\
Amazon2018-Home-And-Kitchen & 776839 & 188643 & 6627987 & 4.12 & 0.00 & 8.53 & 35.14  & 127 & 190 & 147 & 192 & 165 & 200 & 190  \\
Amazon2018-Industrial-And-Scientific & 10715 & 5013 & 70053 & 2.14 & 0.13 & 6.54 & 13.97  & - & - & - & - & - & - & -  \\
Amazon2018-Kindle-Store & 139768 & 98752 & 2217545 & 1.42 & 0.02 & 15.87 & 22.46  & - & 190 & - & 192 & - & - & -  \\
Amazon2018-Luxury-Beauty & 3589 & 1366 & 26784 & 2.63 & 0.55 & 7.46 & 19.61  & - & - & - & - & - & - & -  \\
Amazon2018-Magazine-Subscriptions & 326 & 138 & 2171 & 2.36 & 4.83 & 6.66 & 15.73  & - & - & - & 192 & - & - & -  \\
Amazon2018-Movies-And-TV & 297377 & 59925 & 3281108 & 4.96 & 0.02 & 11.03 & 54.75  & - & 190 & 147 & 192 & - & - & 190  \\
Amazon2018-Musical-Instruments & 27403 & 10449 & 218236 & 2.62 & 0.08 & 7.96 & 20.89  & - & - & - & - & - & - & -  \\
Amazon2018-Office-Products & 101133 & 27500 & 738270 & 3.68 & 0.03 & 7.3 & 26.85  & - & 190 & - & - & - & - & -  \\
Amazon2018-Patio-Lawn-And-Garden & 103102 & 32472 & 744439 & 3.18 & 0.02 & 7.22 & 22.93  & - & 190 & - & - & - & - & -  \\
Amazon2018-Pet-Supplies & 236897 & 42403 & 1948008 & 5.59 & 0.02 & 8.22 & 45.94  & - & 190 & - & - & - & - & -  \\
Amazon2018-Prime-Pantry & 14169 & 4962 & 131058 & 2.86 & 0.19 & 9.25 & 26.41  & - & - & - & - & - & - & -  \\
Amazon2018-Software & 1779 & 729 & 11598 & 2.44 & 0.89 & 6.52 & 15.91  & - & - & - & - & - & - & -  \\
Amazon2018-Sports-And-Outdoors & 331844 & 103911 & 2675299 & 3.19 & 0.01 & 8.06 & 25.75  & - & 190 & 147 & 192 & - & - & -  \\
Amazon2018-Tools-And-Home-Improvement & 240464 & 73153 & 1957540 & 3.29 & 0.01 & 8.14 & 26.76  & - & 190 & - & 192 & - & 200 & -  \\
Amazon2018-Toys-And-Games & 207725 & 78098 & 1754038 & 2.66 & 0.01 & 8.44 & 22.46  & - & 190 & - & - & - & - & -  \\
Amazon2018-Video-Games & 55144 & 17286 & 472857 & 3.19 & 0.05 & 8.57 & 27.35  & - & - & - & - & - & - & -  \\
Anime-Recommendations-Database & 60970 & 8027 & 6314631 & 7.6 & 1.29 & 103.57 & 786.67  & 127 & - & - & 192 & 165 & 200 & 190  \\
Behance & 23724 & 29794 & 687070 & 0.8 & 0.1 & 28.96 & 23.06  & - & 190 & - & - & - & - & -  \\
BookCrossing & 15798 & 38093 & 585579 & 0.41 & 0.1 & 37.07 & 15.37  & - & 190 & - & - & - & - & -  \\
CiaoDVD & 1822 & 2069 & 28144 & 0.88 & 0.75 & 15.45 & 13.6  & - & - & - & - & - & - & -  \\
CiteULike-A & 5543 & 15450 & 205793 & 0.36 & 0.24 & 37.13 & 13.32  & - & - & - & - & - & - & -  \\
CiteULike-T & 4447 & 6811 & 87191 & 0.65 & 0.29 & 19.61 & 12.8  & - & - & - & - & - & - & -  \\
DeliveryHero-SE & 35992 & 18279 & 296647 & 1.97 & 0.05 & 8.24 & 16.23  & - & - & - & - & - & 200 & -  \\
FilmTrust & 1208 & 406 & 31668 & 2.98 & 6.46 & 26.22 & 78.0  & - & - & - & - & - & - & -  \\
FoodComRecipes & 13392 & 32983 & 438576 & 0.41 & 0.1 & 32.75 & 13.3  & - & 190 & - & - & - & - & -  \\
Foursquare-NYC1 & 1261 & 1594 & 9996 & 0.79 & 0.5 & 7.93 & 6.27  & - & - & - & - & - & 200 & -  \\
Foursquare-NYC2 & 1083 & 9989 & 52785 & 0.11 & 0.49 & 48.74 & 5.28  & - & - & - & - & - & 200 & -  \\
Foursquare-Tokyo & 2293 & 15177 & 146998 & 0.15 & 0.42 & 64.11 & 9.69  & - & - & - & - & - & 200 & -  \\
Frappe & 694 & 1435 & 15253 & 0.48 & 1.53 & 21.98 & 10.63  & - & - & - & - & - & 200 & -  \\
GoogleLocal2018 & 65080 & 81193 & 1021304 & 0.8 & 0.02 & 15.69 & 12.58  & - & 190 & - & - & - & - & -  \\
GoogleLocal2021-Alaska & 42808 & 8787 & 680269 & 4.87 & 0.18 & 15.89 & 77.42  & - & - & - & - & - & - & -  \\
GoogleLocal2021-Delaware & 71478 & 11132 & 1150348 & 6.42 & 0.14 & 16.09 & 103.34  & - & - & - & - & - & - & -  \\
GoogleLocal2021-District-Of-Columbia & 74236 & 7747 & 875411 & 9.58 & 0.15 & 11.79 & 113.0  & - & - & - & - & - & - & -  \\
GoogleLocal2021-New-Hampshire & 103256 & 18289 & 1660900 & 5.65 & 0.09 & 16.09 & 90.81  & - & - & - & - & - & - & -  \\
GoogleLocal2021-North-Dakota & 44070 & 8682 & 717137 & 5.08 & 0.19 & 16.27 & 82.6  & - & - & - & - & - & - & -  \\
GoogleLocal2021-Rhode-Island & 67440 & 12134 & 1120757 & 5.56 & 0.14 & 16.62 & 92.37  & - & - & - & - & - & - & -  \\
GoogleLocal2021-South-Dakota & 58439 & 10326 & 893541 & 5.66 & 0.15 & 15.29 & 86.53  & - & - & - & - & - & - & -  \\
GoogleLocal2021-Vermont & 33406 & 7641 & 461690 & 4.37 & 0.18 & 13.82 & 60.42  & - & - & - & - & - & - & -  \\
GoogleLocal2021-Wyoming & 44743 & 8287 & 609316 & 5.4 & 0.16 & 13.62 & 73.53  & - & - & - & - & - & - & -  \\
Hetrec-LastFM & 1859 & 2823 & 71355 & 0.66 & 1.36 & 38.38 & 25.28  & - & - & - & - & - & - & -  \\
Jester-4 & 4101 & 136 & 98559 & 30.15 & 17.67 & 24.03 & 724.7  & - & - & - & - & - & - & -  \\
KGRec-Music & 5199 & 8640 & 751531 & 0.6 & 1.67 & 144.55 & 86.98  & - & - & - & - & - & - & -  \\
LearnFromSets & 854 & 8870 & 450123 & 0.1 & 5.94 & 527.08 & 50.75  & - & - & - & - & - & - & -  \\
Librarything & 27858 & 55542 & 990968 & 0.5 & 0.06 & 35.57 & 17.84  & - & 190 & - & - & - & - & -  \\
MarketBias-ModCloth & 2628 & 763 & 39652 & 3.44 & 1.98 & 15.09 & 51.97  & - & - & - & - & - & - & -  \\
ModCloth-Clothing-Fit & 2238 & 301 & 15799 & 7.44 & 2.35 & 7.06 & 52.49  & - & - & - & - & - & - & -  \\
MovieLens-100K & 943 & 1349 & 99287 & 0.7 & 7.8 & 105.29 & 73.6  & - & - & - & - & - & - & -  \\
MovieLens-1M & 6040 & 3416 & 999611 & 1.77 & 4.84 & 165.5 & 292.63  & - & - & - & - & - & - & -  \\
MovieLens-Latest-Small & 610 & 3650 & 90274 & 0.17 & 4.05 & 147.99 & 24.73  & - & - & - & - & - & - & -  \\
MovieTweetings & 23421 & 11888 & 802784 & 1.97 & 0.29 & 34.28 & 67.53  & - & - & - & - & - & - & -  \\
Myket-Android & 10000 & 7863 & 545817 & 1.27 & 0.69 & 54.58 & 69.42  & - & - & - & - & - & 200 & -  \\
Personality & 1819 & 14885 & 984567 & 0.12 & 3.64 & 541.27 & 66.14  & - & - & - & - & - & - & -  \\
RentTheRunway & 4931 & 3171 & 42481 & 1.56 & 0.27 & 8.62 & 13.4  & - & - & - & - & - & - & -  \\
Retailrocket & 60112 & 34185 & 387914 & 1.76 & 0.02 & 6.45 & 11.35  & - & 190 & - & 192 & - & 200 & -  \\
Steam-Australian-Reviews & 2414 & 576 & 15307 & 4.19 & 1.1 & 6.34 & 26.57  & - & - & - & - & - & - & -  \\
WikiLens & 241 & 1594 & 20539 & 0.15 & 5.35 & 85.22 & 12.89  & - & - & - & - & - & - & -  \\
Yoochoose & 43135 & 4659 & 245210 & 9.26 & 0.12 & 5.68 & 52.63  & - & - & - & - & - & 200 & -  \\
    \bottomrule
  \end{tabular}
\end{table}

\end{document}